\documentclass[prl,aps,reprint,superscriptaddress,nofootinbib]{revtex4-1}

\usepackage[utf8]{inputenc}
\usepackage[english]{babel}
\usepackage{amssymb}
\usepackage{amsmath}
\usepackage{amsfonts}
\usepackage{mathtools}
\usepackage{graphicx}
\usepackage{float}
\usepackage{dsfont} 
\usepackage{cancel}
\usepackage{enumitem}
\usepackage{appendix}
\usepackage{simplewick}
\usepackage{appendix}
\usepackage{bbm}
\usepackage{xcolor}
\usepackage{hyperref}

\newcommand{\ket}[1]{\left| #1 \right>}
\newcommand{\bra}[1]{\left< #1 \right|}
\newcommand{\ketbra}[1]{\left|#1\right>\left<#1\right|}
\newcommand{\Tr}{\text{Tr}}

\newcommand{\bea}{\begin{eqnarray}}
\newcommand{\eea}{\end{eqnarray}}

\newcommand{\nn}{\nonumber}

\newcommand{\be}{\begin{equation}}
\newcommand{\ee}{\end{equation}}
\newcommand{\Op}{\mathcal{O}}

\begin{document}

\title{Geometry of Krylov Complexity}

\author{Pawel  Caputa}
\affiliation{Faculty of Physics, University of Warsaw, ul. Pasteura 5, 02-093 Warsaw, Poland.}
\author{Javier M. Magan}
\affiliation{David Rittenhouse Laboratory, University of Pennsylvania, 209 S.33rd Street, Philadelphia, PA 19104, USA}
\author{Dimitrios Patramanis}
\affiliation{Faculty of Physics, University of Warsaw, ul. Pasteura 5, 02-093 Warsaw, Poland.}

\begin{abstract}

We develop a geometric approach to operator growth and Krylov complexity in many-body quantum systems governed by symmetries. We start by showing a direct link between a unitary evolution with the Liouvillian and the displacement operator of appropriate generalized coherent states. This connection maps operator growth to a purely classical motion in phase space.  The phase spaces are endowed with a natural information metric. We show that, in this geometry, operator growth is represented by geodesics and Krylov complexity is proportional to a volume. This geometric perspective also provides two novel avenues towards computation of Lanczos coefficients and sheds new light on the origin of their maximal growth. We describe the general idea and analyze it in explicit examples among which we reproduce known results from the Sachdev-Ye-Kitaev model, derive operator growth based on SU(2) and Heisenberg-Weyl symmetries, and generalize the discussion to conformal field theories. Finally, we use techniques from quantum optics to study operator evolution with quantum information tools such as entanglement and Renyi entropies, negativity, fidelity, relative entropy and capacity of entanglement. 
\end{abstract}

\maketitle


\section{I. Introduction} \label{sec:I}

The study of classical and quantum chaos is both an exciting and inherently complicated subject. In the classical regime, there is an accepted definition of chaos, based on the behavior of nearby trajectories in phase space and the sensitivity to initial conditions i.e., the butterfly effect \cite{Lorenz}. For chaotic Hamiltonians, trajectories that started infinitesimally close to each other separate exponentially fast with a characteristic Lyapunov exponent. Difficulties in translating such definitions into quantum mechanics have ended up producing many complementary probes of non-equilibrium systems with signatures of quantum chaos \cite{Haake,Berry1989,Larkin}.

Many recent approaches to quantum chaos, that are in one way or another motivated by holography \cite{Maldacena:1997re}, have focused on analyzing characteristic features of the Heisenberg evolution of quantum operators. The main goal there is to quantify the spread/growth of the initial ``simple" operator into ``complex" operators of the model as time evolves. Sensible measures of this ``operator growth'' or ``operator complexity'' are expected to grow exponentially fast in time allowing the definition of different Lyapunov exponents. A related task is to classify families of many-body models from the perspective of operator growth and specify the necessary as well as sufficient conditions for a holographic gravity dual.

When quantifying the operator growth, we face two types of problems: conceptual and technical. Conceptually, the notion of operator size is, to a large extent, arbitrary and it seems unlikely that a single universal quantity will play this role. Indeed, up to date, various ``witnesses" of the growth have been employed. Among them stand out the out-of-time-ordered correlators (OTOC) appearing when computing variances of commutators \cite{Larkin,Shenker:2013pqa,Shenker:2013yza,kitaev,Maldacena:2015waa}. Their analysis culminated in \cite{Maldacena:2015waa}, where a universal upper bound on the Lyapunov exponent, saturated by black holes, was found. This approach is closely related to the eigenstate thermalisation hypothesis \cite{Murthy:2019fgs} and to scrambling \cite{susskind,Barbon1,Lashkari:2011yi,Barbon3,Shenker:2013pqa,freeblack}. Later, a natural definition of operator size in the Sachdev-Ye-Kitaev (SYK) model  \cite{sachdev,kitaev} was analyzed in \cite{Roberts:2018mnp,Qi:2018bje}. More recently, quantum information definitions were employed in \cite{Kudler-Flam:2020yml,MacCormack:2020auw,Kudler-Flam:2019wtv} and a broader approach based on the GNS construction was described in \cite{Magan:2020iac}. Finally, the idea of Krylov complexity, that will be our main topic, was put forward in \cite{Parker:2018yvk}. In all these approaches, various notions of size were shown to evolve exponentially fast and Lyapunov-type exponents were extracted. On the technical side, we need to solve the Heisenberg dynamics in the models of interest. Unfortunately, solving the evolution equation is out of reach in most chaotic quantum systems and thus we are only able to perform numerics in specific models.

In what follows, we will focus on the approach to operator growth and Krylov complexity proposed in \cite{Parker:2018yvk}. This work is rooted in the Lanczos algorithm approach to many-body dynamics \cite{Lanczosbook} (that we review below) used  as a systematic way of constructing a basis in the space of complex operators (Krylov basis). In the same framework, authors introduced a notion of operator complexity,  dubbed ``Krylov complexity'', that can be computed in many-body systems including quantum field theories. In addition, using explicit numerical and analytical examples, the authors put forward a ``universal operator growth hypothesis", arguing that in maximally chaotic systems,  Krylov complexity grows at most exponentially fast with a characteristic Lyapunov exponent. Moreover, they pointed that the operator growth hypothesis might ultimately lead to a new physical proof and understanding of the chaos bound, see also \cite{Murthy:2019fgs}. 

Various aspects of this hypothesis were already investigated in \cite{Rabinovici:2020ryf,Barbon:2019wsy,Jian:2020qpp,Yin:2020oze,Magan:2020iac,Dymarsky:2019elm,Dymarsky:2021bjq,Carrega:2020jrk,Kar:2021nbm,Jian:2020qpp,vonKeyserlingk:2017dyr,Nahum:2017yvy} and Krylov complexity became a good candidate for a universal notion of complexity in interacting quantum field theories. Nevertheless, its physical as well as the operational meaning remain mysterious. On the same footing, the relation to more established notions of complexity is an open problem. On the other hand, despite the relatively unambiguous  definition, computing Krylov complexity requires numerics and understanding its universal features becomes very complicated. These conceptual and technical drawbacks are our main motivations to explore and develop it further in this work.

To make progress, it will be fruitful to focus on certain classes of chaotic models such as those appearing in the context of the AdS/CFT correspondence \cite{Maldacena:1997re}, where, due to conformal symmetry, many-body quantum states are efficiently described geometrically. Indeed, black holes in holography are often seen as collection of qubits (the so-called ``central dogma") described by Hamiltonians that show signatures of maximal quantum chaos. The SYK model  \cite{sachdev,kitaev} described by two-dimensional Anti-de Sitter ($AdS_2$) gravity is the canonical modern example.  Moreover, the quantum information ``revolution" that started with holographic entanglement entropy  \cite{Ryu:2006bv} and continues with holographic complexity \cite{SusskindQC,SStanford,Brown1,Brown2,Aaronson} brought new intuitions that allow us to connect seemingly unrelated concepts from quantum information and computation to geometry (see e.g. reviews \cite{Rangamani:2016dms,Chen:2021lnq}). For instance, microscopic measures of operator growth and complexity are believed to encode subtle information about near horizon geometries of black holes \cite{Susskind:2018tei,Zhao:2017iul,Magan2018,Lin:2019qwu,Magan:2020iac,Barbon:2020uux,Kar:2021nbm,Haehl:2020zgs,Iliesiu:2021ari,DeBoer:2019kdj}.

In this light, we develop a geometric approach to Krylov complexity. Our work will explore the underlying symmetries controlling the system dynamics, although certain observations will be more general. We will be led to the field of generalized coherent states and their associated information geometry. This geometrization will clarify the definition of the operator complexity from a physical standpoint.  More concretely, we will find a precise interpretation of the Krylov complexity as a volume in the information geometry. We will also find the relation between the symmetry algebra governing the operator growth and isometries of this geometry.  At the same time, we will see how this approach simplifies the technical analysis opening new avenues towards the computation of defining aspects of operator growth, such as Lanczos coefficients or Lyapunov exponents in various chaotic and integrable setups. We also notice that the present approach provides a new geometric take on an old field, namely the Lanczos approach to non-equilibrium dynamics, connecting it with the field of generalized coherent states.

This article is organised as follows. In sec II we review the Lanczos algorithm and its recent applications to maximmally chaotic systems. In sec III we describe our main idea that, for symmetry scenarios, the Liouvillian operator can be written in terms of algebra generators as a sum of ``ladder'' operators. This naturally connects with generalized coherents states and their associated geometry. In sec IV we illustrate these ideas in four canonical examples, SL(2,R) (or SU(1,1)), SU(2), Heisenberg-Weyl and 2d CFTs. As highlights, the Lanczos coeffients for SYK, first derived in \cite{Parker:2018yvk} using involved techniques, will acquire a simple and more transparent meaning, and we will determine the geometric roles played by Krylov complexity and the operator wavefunction. In sec V we arrive at the Lanczos coefficients in yet another way, by enforcing the closure of the ladder operator algebra. In sec VI we formulate operator dynamics in terms of a purely classical motion, allowing connections with classical chaos and geometric approaches to complexity. In sec VII, using the two-mode representation of coherent states from quantum optics, we introduce several quantum information tools to probe operator growth: operator entanglement/Renyi entropies, negativity, capacity, fidelity and relative entropy. Finally, in sec VIII we discuss generalizations of Krylov complexity in CFTs and relations to known tools used in discussions of complexity and chaos. Four appendices provide more technical details complementing the discussion in the main part.

\section{ II. Operator Growth and Krylov Complexity} \label{II}
We begin with a brief review of the Lanczos approach \cite{Lanczosbook}  to operator dynamics in many-body systems, leading to the definition of Krylov complexity. We also review previous results for the SYK model that will be reproduced in the following part of the article using our novel approach. Since this topic may not be familiar to a broader audience, we explain it in a slightly pedagogical manner. Readers familiar with the subject may proceed directly to the next section.
\subsection{Operator Growth} 
The Lanczos approach starts with a quantum Hamiltonian $H$ and a time-dependent Heisenberg operator $\Op(t)$ in a given model. The operator can have more labels, such as position, spin, etc., but for the present purposes, only the time-dependence will be explicitly denoted. The evolution of the operator is governed by the Heisenberg equation 
\be
\partial_t\Op(t)=i[H,\Op(t)],
\ee
where $[A,B]=AB-BA$, is the commutator. This equation is formally solved by
\be
\Op(t)=e^{iHt}\,\Op(0)\, e^{-iHt},
\ee
and in what follows, we will denote $\Op(0)=\Op$. The previous expression can be expanded in a formal power series in $t$ as
\be\label{exp}
\Op(t)=\sum^\infty_{n=0}\frac{(it)^n}{n!}\tilde{\Op}_n,
\ee
where $\tilde{\Op}_n$ are nested commutators of $\Op$ with the Hamiltonian
\be
\tilde{\Op}_0=\Op,\quad \tilde{\Op}_1=[H,\Op],\quad \tilde{\Op}_2=[H,[H,\Op]],...
\ee
Knowing the result of these commutators is equivalent to solving the operator dynamics. Unfortunately, this is rarely the case in generic physical systems.

Despite this technical obstruction, we would like to have a notion of growth or complexity of the Heisenberg operator as a function of time. Intuitively, if the Hamiltonian governing the dynamics is sufficiently ``chaotic'', even if we start from a ``simple" operator $\Op$, the result of these commutators will be given by increasingly complex operators. In other words, the more ``chaotic" the Hamiltonian $H$, the faster the operator $\Op$ will mix with other operators of the theory. The main objective is then to quantify such a mixing in a precise manner.
\subsection{Lanczos Algorithm and Krylov Basis} 
In order to sharpen the previous intuitions it will be useful to switch to a better suited formalism and define the Liouvillian super-operator $\mathcal{L}$ (see e.g. \cite{Lanczosbook}) as
\be
\mathcal{L}=[H,\cdot],\qquad \Op(t)\equiv e^{i\mathcal{L}t}\Op,
\ee
and by super-operator we just mean a linear map in the space of operators of the theory. In this language, the operators $\tilde{\Op}_n$ in \eqref{exp} are results of the repeated action of the Liouvillian  $\mathcal{L}$ on $\Op$ such that  $\tilde{\Op}_n\equiv\mathcal{L}^n\Op$.

This view suggests interpreting \eqref{exp} as an ``operator's wavefunction'', and the Liouvillian $\mathcal{L}$ as a Hamiltonian in the Schrodinger formulation. However, we cannot qualify the coefficients of $t^n$ associated with operators $\tilde{\Op}_n$ as ``amplitudes''. One transparent reason is that the sum of their modulus squared is not conserved in time. The precise reason though is that to use the operator algebra as a Hilbert space (in which we expand vectors unambiguously in an orthonormal basis), we need to introduce an inner product. The choice of such an inner product is one of the ambiguities (features) of this approach. In this work, we will follow the most canonical one used in the physics literature.

More concretely, associating $|\mathcal{O})$ with the Hilbert space vector corresponding to operator $\mathcal{O}$, the following family of inner products was described in \cite{Lanczosbook} 
\be
(A|B)^g_\beta=\int^\beta_0 g(\lambda)\,\langle e^{\lambda H}A^\dagger e^{-\lambda H}B \rangle_\beta\, d\lambda. \label{IPGen}
\ee
In this formula, the bracket $\langle\rangle_\beta$ denotes the thermal expectation value 
\be
\langle A\rangle_\beta=\frac{1}{Z}\Tr\left(e^{-\beta H}A\right),\qquad Z=\Tr\left(e^{-\beta H}\right).
\ee
Also, for this definition to be a proper inner-product, $g(\lambda)$ has to satisfy the following conditions
\be
g(\lambda)\ge 0,\quad g(\beta-\lambda)=g(\lambda),\quad \frac{1}{\beta}\int^\beta_0d\lambda g(\lambda)=1.
\ee
In this work, following \cite{Parker:2018yvk}, we will mainly focus on the Wightman inner product
\be
(A|B)=\langle e^{ H\beta/2}A^\dagger e^{-H\beta/2}B \rangle_\beta,
\ee
which corresponds to $g(\lambda)=\delta(\lambda-\beta/2)$. This is a physical choice that amounts to taking the expectation value of the operators in the thermofield double state, with operators  $A$ and $B$ inserted in the two different copies. In any case, once the dynamics is solved for one specific choice of inner product, the behaviour associated with other choices can be found (see e.g. App A in \cite{Magan:2020iac}).

Once we have chosen an inner product, the arbitrary choice of basis in which to expand our evolving operator does not affect the physics of the problem. However, some choices are more convenient than others. Here we will follow the Lanczos approach to non-equilibrium dynamics, which uses the canonical basis generated by the $\vert\tilde{\Op}_n)$. More precisely, starting from $\vert\tilde{\Op}_n)$ and using the Gram–Schmidt orthogonalization procedure we arrive at an orthonormal basis, known as the Krylov basis $|\Op_n)$. In a certain precise sense, this is the ``optimal" choice since the operators $\vert\tilde{\Op}_n)$ are the only ones appearing in \eqref{exp}.

The Krylov basis is defined recursively using the following algorithm (also known as Lanczos algorithm). We start by noticing that the first two operators in $\vert\tilde{\Op}_n)$ are always orthogonal with respect to the previous inner products~(\ref{IPGen}). Therefore we can directly include them in our basis
\be
|\Op_0):=|\tilde{\Op}_0)=|\Op),\qquad |\Op_1):=b^{-1}_1\mathcal{L}|\tilde{\Op}_0),
\ee
where $b_1=(\tilde{\Op}_0\mathcal{L}|\mathcal{L}\tilde{\Op}_0)^{1/2}$ normalizes the vector. The next states are constructed iteratively by first computing 
\be
|A_n)=\mathcal{L}|\Op_{n-1})-b_{n-1}|\Op_{n-2}),\label{AnDef}
\ee 
and then normalizing
\be
|\Op_n)=b^{-1}_n|A_n),\qquad b_n=(A_n|A_n)^{1/2}.
\ee
This way, we arrive at an orthonormal basis $( \Op_n| \Op_m)=\delta_{n,m}$ that has been generated by the set $\{\mathcal{L}^n\Op\}$. We can now use it to expand any element of this set and the evolving operator $|\Op (t))$. Notice that in addition to the Krylov basis states $|\Op_n)$, this algorithm yields the so-called Lanczos coefficients $b_n$. Finding these coefficients for the system under consideration amounts to solving for the dynamics and it is one of the technical challenges in this approach, see \cite{Lanczosbook}. Let us also point that the above algorithm can be generalized to include diagonal terms in the Liouvillian (see e.g. Appendix A).

We now expand the time-dependent operator in the Krylov basis as
\be
|\Op(t))=\sum_n i^n\varphi_n(t)|\Op_n)\;.\label{OTKrB}
\ee
In this expansion, the amplitudes $\varphi_n(t)$ turn out to be real. Generally, their modulus squared defines probabilities whose sum is conserved in time
\be
\sum_{n}|\varphi_n(t)|^2\equiv\sum_{n}p_n(t)=1.
\ee
These amplitudes are determined by solving a ``Schrodinger equation", that descends from the original Heisenberg equation satisfied by $\Op(t)$. To derive this equation, notice that the previously defined Liouvillian $\mathcal{L}$ plays the role of the Hamiltonian in the new Hilbert space spanned by the Krylov basis $|\Op_n)$. In particular, the state representing $\Op(t)$ is given by
\be
|\Op(t))=e^{i\mathcal{L}t}|\Op).\label{eqL}
\ee
Computing the time derivative
\bea
\partial_t|\Op(t))&=&i\mathcal{L}|\Op(t)),\label{EQPDrO}
\eea
or equivalently, using \eqref{OTKrB} we arrive at
\be
\partial_t|\Op(t))=\sum_n i^n\partial_t\varphi_n(t)|\Op_n).\label{PDOpSF}
\ee
Next, from the Lanczos algorithm \eqref{AnDef}, we find the action of the Liouvillian on the Krylov basis vectors 
\be
\mathcal{L}|\Op_n)=b_n|\Op_{n-1})+b_{n+1}|\Op_{n+1}).\label{LinKB}
\ee
From this expression it is clear that the Liovillian is tridiagonal in the Krylov basis (generally we may have a diagonal term in \eqref{LinKB}). This fact will play an important role in the following sections. Applying this to \eqref{EQPDrO} and shifting the summation appropriately, we derive
\be
\partial_t|\Op(t))=\sum_{n}i^n\left(b_n\varphi_{n-1}(t)-b_{n+1}\varphi_{n+1}(t)\right)|\Op_n).\label{PDOprMEXP}
\ee
Comparing the coefficients of \eqref{PDOpSF} and \eqref{PDOprMEXP}, we arrive at the discrete Schrodinger equation determining the time evolution of the amplitudes $\varphi_n(t)$
\be\label{SchrodingerEq}
\partial_t\varphi_n(t)=b_n\varphi_{n-1}(t)-b_{n+1}\varphi_{n+1}(t)\;.
\ee
With this equation, once we derive the Lanczos coefficients $b_n$, we can solve for the amplitudes $\varphi_n(t)$ with initial condition $\varphi_n(0)=\delta_{n0}$ and determine the operator wavefunction \eqref{OTKrB}. The operator's wavefunction then completely determines the growth of the operator that, as we will describe below, can be measured using tools of quantum mechanics, quantum information, or quantum complexity.

Before we discuss operator's complexity, we note that a very special role in the Krylov approach is played by the so-called auto-correlation function   
\be 
C(t)\equiv (\Op(t)|\Op)=\varphi_0(t)\;.
\ee
Indeed, as reviewed in \cite{Parker:2018yvk}, starting from $C(t)$ and/or its appropriate transforms we can obtain the Lanczos coefficients $b_n$ and operator wavefunction. In this work, it will be more instructive to develop our physical understanding of the Liouvillian instead. This will allow us to easily extract both $C(t)$ and $b_n$.
\subsection{Krylov Complexity} 
We now describe how to quantify operator complexity in this framework. Using physical intuition,  we can first interpret the dynamics in equation~(\ref{SchrodingerEq}) as that of a particle moving on a one-dimensional chain, where the sites with label $n$ are in one-to-one correspondence with the Krylov basis vectors (see also \cite{Dymarsky:2019elm} for a Toda chain perspective). This suggests a natural measure of operator complexity, dubbed Krylov complexity  \cite{Parker:2018yvk}, defined to be the average position in the chain
\be 
K_{\mathcal{O}}\equiv \sum_{n} n \,p_n(t)=\sum_{n} n \,\vert \varphi_n(t)\vert^2\;.
\ee
Formally, this quantity can be written as the expectation value in the evolving state $|\Op(t))$ of the following ``Krylov complexity operator''
\be 
\hat{K}_{\mathcal{O}}=\sum_{n} n |\Op_n)(\Op_n\vert\;,\label{KOper}
\ee
such that Krylov complexity reads
\be \label{kex}
K_{\mathcal{O}}=(\Op(t)|\hat{K}_{\mathcal{O}}|\Op(t))\;.
\ee
Intuitively, this position operator \eqref{KOper} in the chain can also be interpreted as a ``number operator". Unlike the Liouvillian, it is diagonal in the Krylov basis.

Clearly, as with the choice of the inner product, there is a certain ambiguity in this definition of operator complexity. Indeed, several definitions of operator complexities that have appeared in the literature can always be written in such a way, see \cite{Parker:2018yvk,Magan:2020iac}. However, as we will see in this work, this ``minimal" choice acquires a simple geometric interpretation.

The recent interest in the Krylov approach to operator complexity has various origins. First, modulo simple physical assumptions, it is a well defined and concrete approach, potentially applicable to QFTs. These features make it appealing from the point of view of holography. Second, based on various explicit numerical as well as analytical examples, \cite{Parker:2018yvk} conjectured a maximal possible growth of Lanczos coefficients in quantum systems, namely a linear growth:
\be\label{bounb}
b_n\leq \alpha n+\gamma+O(1),
\ee
where $\alpha$ is the operator growth rate and $\gamma$ is a non-universal constant that depends on the details of the operator. In particular, for this type of Lanczos coefficients, i.e., systems saturating the bound, the Krylov complexity grows exponentially fast with an exponent given by $\lambda=2\alpha$. In several examples, some of which will be described below, at finite temperature $T=1/\beta$ one arrives at $\alpha=\pi/\beta$, and this was conjectured to bound the Lyapunov exponent, as defined by out-of-time ordered correlation functions \cite{Maldacena:2015waa}.
\subsection{SYK example} 
As the key example of the behaviour~(\ref{bounb}), the SYK model \cite{sachdev,kitaev}, which is a modern playground for quantum chaos \cite{kitaev,Maldacena:2015waa}, was analyzed in \cite{Parker:2018yvk}. 
The SYK model \cite{sachdev,kitaev} is a model of $N$ Majorana fermions interacting with all-to-all random couplings. For random $q$-body interactions, the Hamiltonian is of the form
\begin{equation}\label{SYK}
H=i^{q/2} \sum\limits_{1\leq i_1<i_2< \cdots <i_q\leq N}J_{i_1 i_2 \cdots i_q}\psi_{i_1}\psi_{i_2}\cdots\psi_{i_q}\;,
\end{equation} 
This model has been at the center of attention for the past years for several important reasons, namely exact solvability at large $N$, conformal phase at low energies, and maximal chaos in the sense of \cite{Maldacena:2015waa}.

Operator growth for this system was considered in \cite{Roberts:2018mnp}, using a natural notion of growth arising from the exact Majorana fermion formulation of the model. An advantage of such an approach is that it was naturally related to out-of-time ordered correlation functions, see also \cite{Qi:2018bje}. A disadvantage is that such a definition does not seem to find a natural extension to higher dimensions and QFTs.

Operator growth for this system was also reconsidered in \cite{Parker:2018yvk} using the Lanczos approach. As explained above, the starting point of this approach can be taken to be the autocorrelation function. For SYK at low temperatures this is
\be\label{autosyk}
C(t)=\cosh^{-\eta}\left( \frac{\pi t}{\beta}\right).
\ee
In this case, the Lanczos coefficients can be obtained analytically \cite{Parker:2018yvk} (see also \cite{Dymarsky:2019elm}) and are given by
\be
b_n=\frac{\pi }{\beta}\sqrt{n(\eta+n-1)}\label{Ubn}\;.
\ee
The operator wavefunction can then be found by solving \eqref{SchrodingerEq} and reads
\be
 \varphi_n(t)=\sqrt{\frac{\Gamma(\eta+n)}{n!\Gamma(\eta)}}\frac{\tanh^n(\alpha t)}{\cosh^\eta(\alpha t)}.\label{Upn}
\ee
The probabilities $p_n(t)=|\varphi_n(t)|^2$ from this solution correspond to the negative binomial distribution. The evolution of these probabilities depicts a one-dimensional diffusion process over the Krylov basis. The time evolution of the mean position in this chain, or equivalently the evolution of Krylov complexity, is of exponential type. It is controlled by the maximal Lyapunov exponent $\lambda=2\pi/\beta$. More explicitly
\be
K_\Op=\eta \sinh^2(\alpha t)\sim \frac{\eta}{4}e^{2\alpha t}=e^{2\alpha\left(t-\frac{1}{2\alpha}\log\left(\frac{4}{\eta}\right)\right)}\;,
\ee
where we have written the coefficient of the exponent in an analogous way to the scrambling time in the OTOC. Observe that, while the exponential growth is ``more universal'' than the usual Lyapunov growth (it does not receive stringy corrections for example in the context of holography), the ``scrambling time'' for a given operator is by construction less universal. Nevertheless, it depends on the scaling dimension of the initial perturbation and may also be a good probe for the operator growth.

Before moving forward we want to make a couple of remarks. First, from a technical standpoint, the derivation of the operator wavefunctions in both \cite{Roberts:2018mnp} and \cite{Parker:2018yvk} is quite involved.  This feature makes it difficult to extrapolate to other systems, in particular to higher dimensions. On the other hand, readers familiar with the SYK model and the arguments that lead to the derivation of the correlator \eqref{autosyk} (using large-N techniques, see \cite{sachdev,kitaev}) may recall it was the conformal symmetry appearing in the low energy Schwinger-Dyson equations that was responsible for the form of this two-point function. In other words, the fermions behave as primaries transforming in specific representations of the SL(2,R) algebra. In particular, for the $q$-body interaction, the associated scaling dimension is $h=1/q$. We might expect a deeper and simpler understanding of operator dynamics and wavefunction when such a feature is included in the analysis.

Second, from a more holographic standpoint, the relation between Krylov complexity and the actual physics of the problem is far from clear. In the light of recent discussions on near horizon symmetries in black hole physics and their potential connections with operator complexity \cite{Susskind:2018tei,Magan2018,Lin:2019qwu,Magan:2020iac,Barbon:2020uux,Kar:2021nbm}, we would like to have a better understanding of the Krylov complexity operator. \\
In the following sections, we will explore a geometric avenue towards both problems, which more broadly can be seen as a new perspective on the Lanczos approach.

\section{III. Liouvillian and symmetry: General Idea} \label{sec:III}
In this section, we describe a general paradigm that we will follow through the rest of the article. The main idea is simple yet powerful, and we describe it below. From the zoo of complicated quantum systems, we focus our attention on models governed by symmetry. By this, we mean systems for which the Liouvillian operator belongs to the Lie algebra of a given symmetry group. In the context of the usual Shrodinger evolution, this is quite a common lore. For example, in QFT or CFT the Hamiltonian belongs to the Lie algebra of the Poincar\'e group or the conformal group, respectively. This idea is old and well explored in Hamiltonian dynamics (see e.g. review \cite{RevModPhys.62.867}). Here, we import it to the physics of operator evolution, instead of state evolution, where the Liouvillian plays the role of the Hamiltonian in the Krylov basis. 

With symmetries in mind, our key observation is that the action of the Liouvillian on the Krylov basis \eqref{LinKB} can be interpreted as the action of the sum of abstract ``raising''  and ``lowering'' ladder operators $L_+$ and $L_-$, namely
\be \label{liug}
\mathcal{L}=\alpha\,(L_+ +L_-)\;.
\ee
In this expression, $\alpha$ is a proportionality factor, not fixed by symmetry. It will depend on the details of the physical setup, such as the choice of the inner-product, etc. Its meaning will become clearer in the examples below.\\
With such Liouvillians, the Krylov basis states will naturally furnish representations of the appropriate symmetry group. This is again analogous to relativistic QFT or CFT, where states are organized through representations of the Poincar\'e or conformal group. The only difference here is that we apply such a structure to operator dynamics on the Krylov basis.

In the light of symmetry, the previously described quantities associated with the Lanczos approach take a more transparent meaning. First, since the action of the ladder operators in a certain representation is fixed by the symmetry group, this approach allows us to read off the Lanczos coefficients immediately. More precisely, they are simply determined from the action of ladder operators in the Krylov basis 
\bea 
\alpha L_+|\Op_n)=b_{n+1}|\Op_{n+1}),\quad \alpha L_-|\Op_n)=b_{n}|\Op_{n-1}).\,\,~
\eea
We will also see that, under certain conditions, the Lie group approach leads to quadratic algebraic equations for Lanczos coefficients. This will ensure that, at least in our examples, they will not grow faster than $n$, in agreement with the maximal operator growth hypothesis \cite{Parker:2018yvk}. 

Moreover, the above paradigm allows us to make a powerful connection with generalized coherent states \cite{coherent1,coherent2,gazeau2009}. This comes from the fact that the Liouvillian time evolution in the Krylov basis with \eqref{liug} can be seen as a particular instance of a generalized displacement operator $D(\xi)$ for a Lie group. These displacements operators typically take the form 
\be 
D(\xi)\equiv e^{\xi L_+-\bar{\xi} L_-}\;,
\ee
for some complex $\xi$, its conjugate $\bar{\xi}$ and the same abstract ladder operators $L_\pm$. We will make all these formulas precise when analyzing specific examples in the next section. The coherent state can now be written as the action of the displacement operator on some reference state $\ket{\Psi_0}$, usually chosen to be the highest weight state of the representation. It is clear that unitary time evolution, as generated by the Liouvillian  \eqref{liug}, is just a displacement operator with $\xi=i\alpha\,t$. In other words, we can interpret the operator dynamics and its growth in the Krylov basis as a trajectory through the Hilbert space of coherent states. This way, after associating $\ket{\Psi_0}$ with our initial operator $|\Op)$, and expanding the coherent states in an orthonormal basis, we will be able to read off the amplitudes $\varphi_n(t)$ and the Krylov basis vectors $|\Op_n)$.

The link with coherent states further allows us to geometrize Krylov complexity. This formulation is rooted in the well-known connection between coherent states and information metric (Fubini-Study metric) on the Hilbert space, abstractly defined for the coherent state $\ket{z}$ as
\be 
ds^2_{FS}=\langle dz|dz\rangle-\langle dz|z\rangle\langle z|dz\rangle.
\ee
This metric is also associated with the coadjoint orbit of the relevant group (see e.g. \cite{RevModPhys.62.867}). As we will see, the Krylov complexity will be universally proportional to the ``Volume" in this geometry.  In addition, both the Liouvillian $\mathcal{L}$ as well as the Krylov complexity operator $\hat{K}_\Op$ can be related to isometry generators in these information geometries. Indeed they form a ``complexity algebra'' isomorphic to the algebra of isometries and we will show how it determines Lanczos coefficients.

Finally, the association of the coherent state complex label $\xi$ with real-time suggests that we are secretly discussing a classical motion in phase space. This interpretation is indeed correct and it paves a way towards understanding the relations between Krylov complexity and circuit complexity.
\section{IV. Liouvillian and symmetry: Examples} \label{sec:IV}
In this section, we analyze explicit examples of the general idea above. From a physical perspective, the most interesting one is that of SL(2,R) and its generalizations to Conformal Field Theories (CFT). These have applications to classical and quantum chaos and the physics of black holes. We will also discuss the examples of SU(2) and the Heisenberg-Weyl group, which will help us gain more intuition about the relation between Krylov complexity, group theory, and geometry.
\subsection{Example I: SL(2,R)  } 
The first example is operator evolution governed by SL(2,R). In this case, we will re-derive the SYK results of \cite{Parker:2018yvk} using the above general paradigm.\\ We start from the commutation relations for the SL(2,R) algebra 
\be
[L_0,L_{\pm 1}]=\mp L_{\pm 1},\qquad [L_1,L_{-1}]=2L_0,\label{SL2R}
\ee
and consider a discrete series representation labeled by a positive integer $h$. This representation is typically expanded by orthonormal vectors $\ket{h,n}$, for $n$ a non-negative integer, satisfying $\bra{h,m}h,n\rangle=\delta_{n,m}$. The basis vectors are eigenstates of the $L_0$ operator as well as the Casimir operator $C_2=L^2_0-\frac{1}{2}(L_{-1}L_1+L_{1}L_{-1})$ with eigenvalue $h(h-1)$. The full action of the SL(2,R) generators in this basis is given by
\bea
L_0\ket{h,n}&=&(h+n)\ket{h,n},\nn\\
L_{-1}\ket{h,n}&=&\sqrt{(n+1)(2h+n)}\ket{h,n+1},\nn\\
L_1\ket{h,n}&=&\sqrt{n(2h+n-1)}\ket{h,n-1},\label{L1On}
\eea
which in particular implies that
\be\label{l1n}
\ket{h,n}=\sqrt{\frac{\Gamma(2h)}{n!\Gamma(2h+n)}}L^n_{-1}\ket{h}\;.
\ee
The same Hilbert space can be also expanded by means of generalized coherent states, see \cite{coherent2}, that are defined by using the displacement operator
\be
\ket{z,h}\equiv D(\xi)\ket{h},\qquad D(\xi)=e^{\xi L_{-1}-\bar{\xi} L_1},\label{SLRCS}
\ee
where the relation between the complex variables is
\be\label{zxi}
z=\frac{\xi}{|\xi|}\tanh(|\xi|),\qquad |\xi|=\sqrt{\xi \bar{\xi}}.
\ee
It is useful to introduce polar coordinates $\xi=\frac{1}{2}\rho e^{i\phi}$, such that $z$ parametrizes the unit disc
\be
z=\tanh\left(\frac{\rho}{2}\right)e^{i\phi},\qquad |z|<1.
\ee
Using the action of the SL(2,R) generators on the primary state, in particular relation~(\ref{l1n}), we can write these so-called SU(1,1) Perelomov coherent states more explicitly as
\be
\ket{z,h}=(1-|z|^2)^h\sum^\infty_{n=0}z^n\sqrt{\frac{\Gamma(2h+n)}{n!\Gamma(2h)}}\ket{h,n}.\label{PCS}
\ee
Now we will follow the general paradigm described in the previous section. First, from \eqref{L1On}, we note that $L_{-1}$ is playing the role of the abstract raising operator $L_+$ and $L_{1}$ of the lowering operator $L_-$. This way, the Liouvillian governing the SL(2,R) operator dynamics in the Krylov basis is given by
\be
\mathcal{L}=\alpha\,(L_{-1}+L_{1})\;.\label{LLSU2R}
\ee
As reviewed above, the operator wavefunction \eqref{OTKrB} is obtained by applying the unitary evolution with $\mathcal{L}$, so that
\be
|\Op(t))=e^{i\alpha(L_{-1}+L_1)t}\ket{h}.
\ee
Returning to the definition of the coherent state \eqref{SLRCS}, we make the key observation that our operator's wavefunction is nothing but the Perelomov coherent state with $\xi=i\alpha t$. More explicitly we have the relation
\be
|\Op(t))=\ket{z=i\tanh(\alpha t),h=\eta/2},
\ee
as well as the identification between the Krylov basis and the basis vectors associated with representation $h$ of the  SL(2,R) group
\be
|\Op)=\ket{h},\qquad |\Op_n)=\ket{h,n}.
\ee
Arguably the most elegant consequence of this map is the fact that from the action of the ladder operators  \eqref{L1On}, we immediately get the Lanczos coefficients 
\be
b_n=\alpha\sqrt{n(2h+n-1)}.
\ee
We can indeed check that the wavefunctions \eqref{Upn} are just coefficients of the coherent state \eqref{PCS} with $z=i\tanh(\alpha t)$ and solve the Schrodinger equation \eqref{SchrodingerEq} with the Lanczos coefficients above. \\
The Krylov complexity is then proportional to the highest weight $h$ and grows exponentially with time, with the Lyapunov exponent $\lambda=2\alpha$
\be
K_\Op=(\Op(t)|n|\Op(t))=2h\sinh^2(\alpha t).
\ee
Moreover, the $n=0$ amplitude is the SYK auto-correlation function \eqref{autosyk} with $\eta=2h$.

A more appropriate interpretation of the identification $\xi=i\alpha t$ is that operator dynamics in this setup is mapped to a particular classical trajectory in the phase space of coherent states. In polar coordinates, this trajectory corresponds to setting $\rho=2\alpha t$ and $\phi=\pi/2$. We will return to this interpretation in section VII.

We now introduce the ``information geometry" associated with generalized coherent states and use it to interpret operator growth and Krylov complexity geometrically. To this end, we recall that in a quantum theory, the space of coherent states has an associated geometry described by the Fubini-Study metric (also dubbed information metric). For our states \eqref{PCS} this becomes the standard metric on the hyperbolic disc. In complex coordinates $(z,\bar{z})$ as well as in $(\rho,\phi)$ it reads
\be
ds^2_{FS}=\frac{2hdzd\bar{z}}{(1-z\bar{z})^2}=\frac{h}{2}\left(d\rho^2+\sinh^2(\rho)d\phi^2\right).\label{FSSU11}
\ee
With this geometry at our disposal, we want to make several comments. Firstly, the identification used to describe the growth, namely $\rho=2\alpha t$ and $\phi=\pi/2$, defines a geodesic in this geometry. In other words, the operator growth process gets mapped to a geodesic motion in a hyperbolic geometry \eqref{FSSU11}. This will be made more precise in section VII. Secondly, we can interpret the Krylov complexity operator as a generator of translations in the $\phi$ direction (an isometry generator). This is seen from the explicit form of the coherent state, and the fact that $-i\partial_\phi$ produces a factor $n$. We will discuss more precisely the relation between the isometries of this information geometry and the complexity algebra generated by the Liouvillian and the Krylov complexity operator in a later section.

\begin{figure}[h!]
  \centering
  \includegraphics[width=6cm]{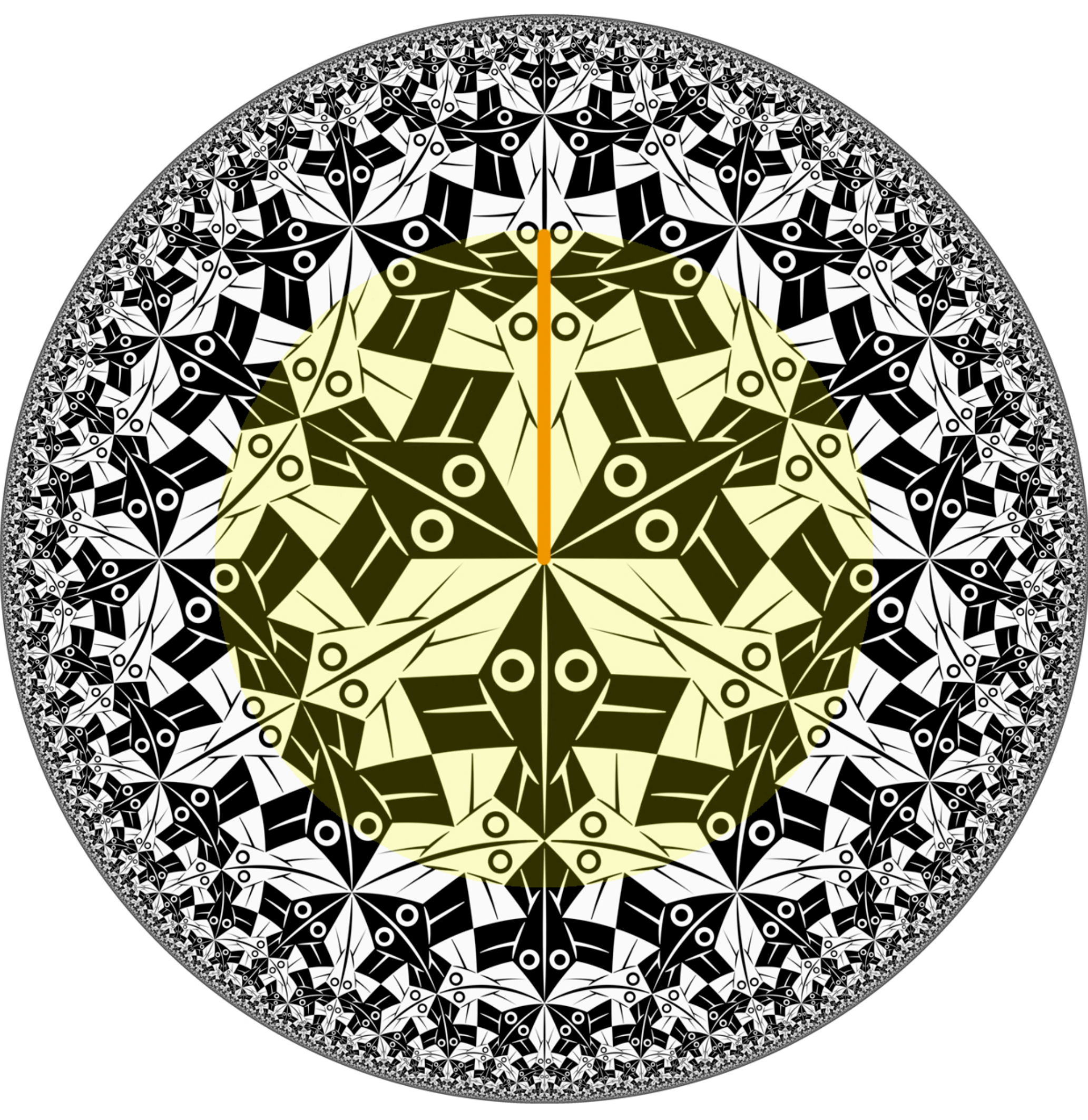}
  \caption{Cartoon of the operator growth and Krylov complexity for SL(2,R). Coherent states allow us to map the operator evolution to a geodesic (in orange) on the hyperbolic disc. Volume (in yellow) of the region enclosed by the particle's position $\rho=2\alpha t$ at $\phi=\pi/2$ is proportional to the Krylov complexity. }
\label{Volume}
\end{figure}

Thirdly, motivated by the recent developments concerning the geometric approach to complexity, we note that the actual Krylov complexity is proportional to the volume enclosed by the geodesic radius $\rho=\alpha t$, i.e., it is proportional to the volume of the region from the origin $\rho=0$ up to $\rho=2\alpha t$ (see Fig. \ref{Volume}). The explicit computation gives
\be
V_t=\int^{2\alpha t}_0d\rho\int^{2\pi}_0d\phi\sqrt{g}=2\pi h\sinh^2(\alpha t)=\pi K_\Op.
\ee
This is one of the main new results of our work. We will show that this relation holds more generally in other examples.

Based on the intuition from Nielsen's approach to circuit complexity, to be described later, one may have naively expected a relation between the geodesic length and complexity. However,  the geodesic distance between two arbitrary points $(\rho_i,\phi_i)$ and $(\rho_f,\phi_f)$ in geometry \eqref{FSSU11} is given by
\be
\cosh(L/l)=\cosh(\rho_f)\cosh(\rho_i)-\cos(\Delta\phi)\sinh(\rho_f)\sinh(\rho_i),
\ee
where the radius of the hyperbolic space is denoted as $l^2=h/2$. This way, if we measure it from the center of the disc $\rho_i=0$, the geodesic length is $L=\rho_f$. For our geodesic motion, we have $\rho_f=\alpha t$, which only grows linearly in $t$. We will also return to this point in a later section, where we will see the more direct relation to Nielsen's complexity.

Last but not least, geometry \eqref{FSSU11} is negatively curved. The Ricci scalar is related to the highest weight state $h$ as
\be
R=-\frac{4}{h},
\ee
and it decreases for large $h$.\\
There have already been several discussions, both in classical and quantum chaos as well as in the complexity literature, about the role of negatively curved information geometry \cite{Brown:2016wib,Lin:2019kpf,Auzzi:2020idm,Flory:2020dja,Basteiro:2021ene}. In the context of black holes, for example, perturbations in the near horizon region can be described in this way, see \cite{Barbon1,Barbon3,Barbon2,Barbon4}. The present example is a precise contribution to this intuition. Indeed, we will see in the following examples that the sign of the curvature is correlated with the nature of Krylov complexity growth.

\subsection{Example II: SU(2) } 
We now analyze the example in which the Liouvillian belongs to the SU(2) algebra. This will give us a more general intuition about the Krylov approach in non-chaotic systems. In particular, we will see the consequences of working with a finite-dimensional Hilbert space and having non-maximal Lanczos coefficients on the Krylov complexity and its geometry.

We start with the familiar SU(2) Lie algebra
\be
[J_i,J_j]=i\epsilon_{ijk}J_k,
\ee
and introduce the ladder operators $J_{\pm}=J_1\pm iJ_2$. Renaming $J_3\rightarrow J_0$ the previous algebra transforms into
\be
[J_0,J_\pm]=\pm J_\pm,\qquad [J_+,J_-]=2J_0.
\ee
Using the ladder operators we can build the usual basis for representation $j=0,\frac{1}{2},1,\cdots$, namely $\vert j,n\rangle$, with $-j\leq n\leq j$. In order to make the connection with operator growth, it will be convenient to re-label the basis vectors as $n\rightarrow j+n$, so that $n=0,...,2j$. This way, the  $2j+1$ orthonormal basis vectors can be written as
\be
\ket{j,-j+n}=\sqrt{\frac{\Gamma(2j-n+1)}{n!\Gamma(2j+1)}}J^{n}_+\ket{j,-j}\;.
\ee
In this basis, the action of the Lie algebra generators is
\bea
J_0\ket{j,-j+n}&=&(-j+n)\ket{j,-j+n},\nn\\
J_+\ket{j,-j+n}&=&\sqrt{(n+1)(2j-n)}\ket{j,-j+n+1},\nn\\
J_-\ket{j,-j+n}&=&\sqrt{n(2j-n+1)}\ket{j,-j+n-1}.\label{Jmrepn}
\eea
As before, we will choose the highest weight state $\ket{j,-j}$, annihilated by $J_-$, as our initial state. Equivalently we could have started from $J_+\ket{j,j}=0$ but we chose to follow the usual convention \cite{coherent2}.

Following previous steps, we build the so-called spin coherent states by applying the displacement operator
\be
\ket{z,j}=D(\xi)\ket{j,-j},\qquad D(\xi)=e^{\xi J_+-\bar{\xi}J_-},
\ee
where now we have the complex coordinate
\be
z=\tan\left(\frac{\theta}{2}\right)e^{i\phi},
\ee
that parametrizes a spherical geometry.

\begin{figure}[t!]
  \centering
  \includegraphics[width=8cm]{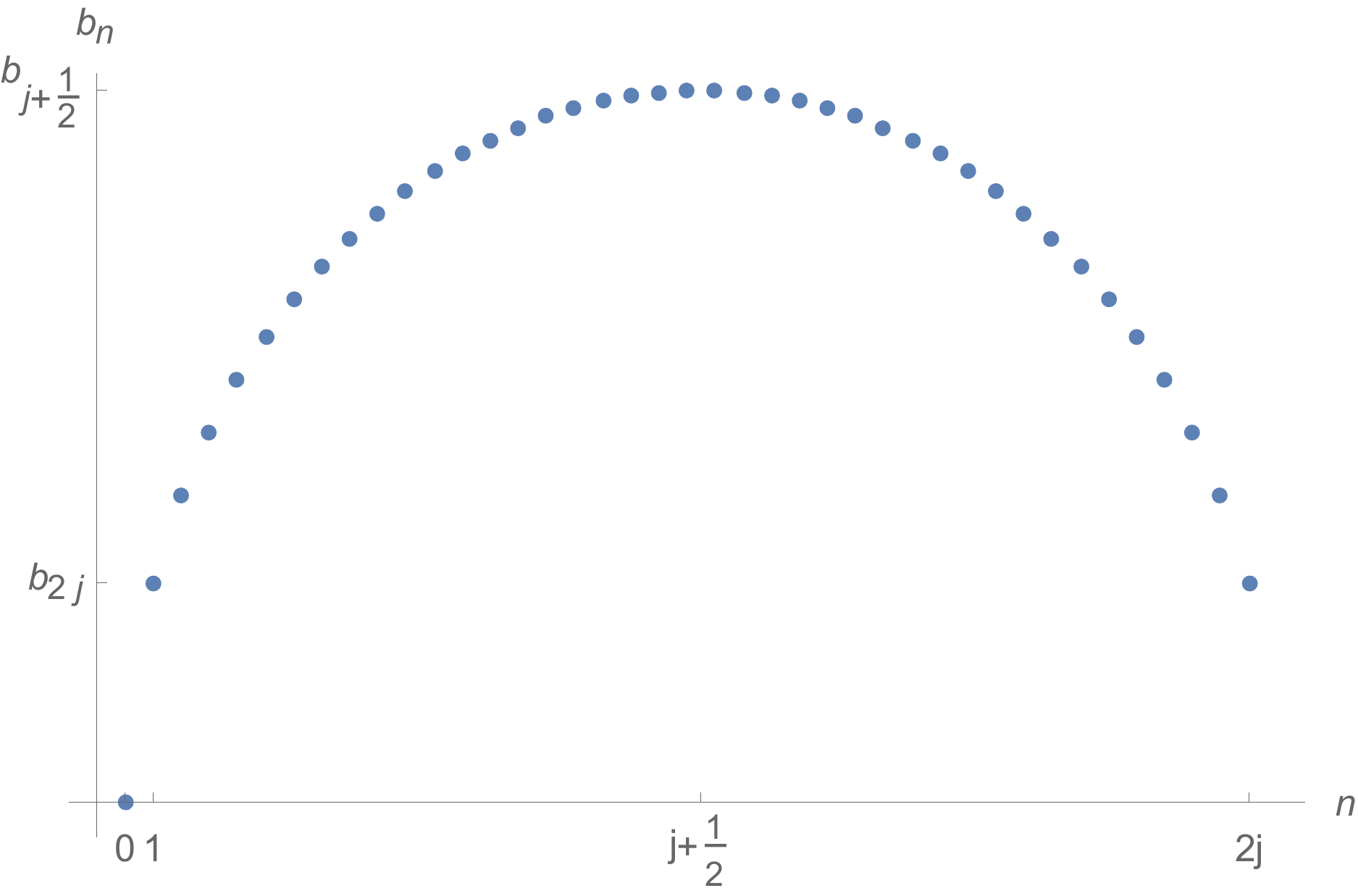}
  \caption{Distribution of the SU(2) Lanczos coefficients. Sample plot for $j=20$.}
\label{LanczosSU(2)Plot}
\end{figure}

\begin{figure}[b!]
  \centering
  \includegraphics[width=8cm]{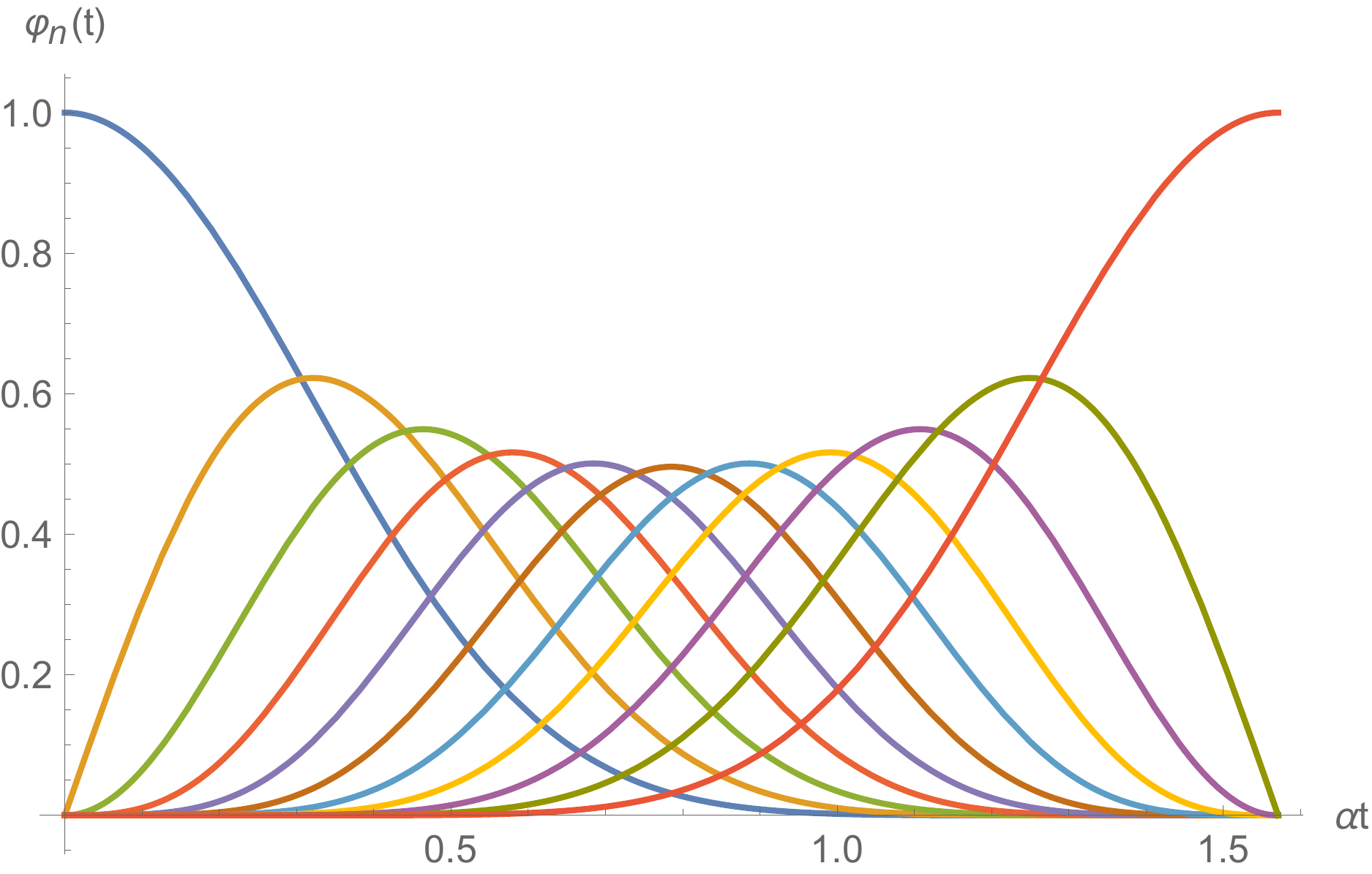}
  \caption{All the 11  wavefunctions $\varphi_n(t)$ for spin $j=5$ plotted between $\alpha t\in(0,\pi/2)$. Different wavefunctions are peaked at later values of $\alpha t$ symmetrically, reflecting the symmetry of $b_n$'s.}
\label{WFSU(2)EPlot}
\end{figure}

More explicitly, the spin coherent states are written in the orthonormal basis as
\be
\ket{z,j}=(1+z\bar{z})^{-j}\sum^{2j}_{n=0}z^n\sqrt{\frac{\Gamma(2j+1)}{n!\Gamma(2j-n+1)}}\ket{j,-j+n}.\label{CSSU2}
\ee
To analyze the operator growth we repeat the same steps as in the previous example. The SU(2) Liouvillian takes the form
\be
\mathcal{L}=\alpha(J_++J_-)\;,
\ee 
and we find finite-dimensional Hilbert spaces, with dimensions $2j+1$, where Krylov basis states are associated with orthonormal vectors in an obvious way
\be
 |\Op_n)=\ket{j,-j+n},\quad n=0,...,2j.
\ee
With this identification, the action of the lowering operator $J_-$ in \eqref{Jmrepn} automatically allows us to read off the $b_n$'s
\be
b_n=\alpha\sqrt{n(2j-n+1)}.\label{bnSU2}
\ee
These Lanczos coefficients grow slower than their SL(2,R) cousins, namely as $\alpha\sqrt{2j\,n}$ up to a maximum value
\be
n_{max}=j+\frac{1}{2},\quad b_{j+\frac{1}{2}}=\alpha\left(j+\frac{1}{2}\right)=\alpha n_{max},
\ee
and then come back down to the final value (see example on Fig. \ref{LanczosSU(2)Plot})
\be
b_{2j}=\alpha\sqrt{2j}=b_1.
\ee
Using the above form of the spin coherent states, we find the Heisenberg operator wavefunction $|\Op(t))$ in the Krylov space by replacing $\theta=2\alpha t$ and $\phi=\pi/2$. More precisely
\be
|\Op(t))=\ket{z=i\tan(\alpha t),j}=e^{i\alpha(J_++J_-)t}\ket{j,-j}.
\ee
The SU(2) wavefunction arising from this identification
\be
\varphi_n(t)=\frac{\tan^n(\alpha t)}{\cos^{-2j}(\alpha t)}\sqrt{\frac{\Gamma(2j+1)}{n!\Gamma(2j-n+1)}},\label{WFSU2}
\ee
satisfies the Schrodinger equation \eqref{SchrodingerEq} with Lanczos coefficients \eqref{bnSU2}. To get intuition about the shape of these functions, we plot the example of $j=5$ in Fig~(\ref{WFSU(2)EPlot}). 
In this SU(2) case, the probabilities $p_n(t)$ form the binomial distribution
\be
p_n(t)=|\varphi_n(t)|^2=\binom{2j}{n}\lambda^{n}(1-\lambda)^{2j-n},
\ee
with $\lambda=\sin(\alpha t)$. The auto-correlation function for SU(2), from which one obtains the return probability, is given by
\be
C(t)=\varphi_0(t)=\frac{1}{\cos^{2(-j)}(\alpha t)}.
\ee
This correlation function appears e.g. when analyzing a free harmonic oscillator at finite temperature. The two-point function in Euclidean time for such an oscillator is (see e.g. \cite{PLaine_2016})
\be 
G(\tau)=\frac{1}{2\omega}\frac{\cosh [(\beta/2-\tau)\omega]}{\sinh (\beta\omega/2)}\;.
\ee
After doing the usual analytic continuation $\tau\rightarrow it$, and then the analytic continuation towards the inner product $t\rightarrow t-i\beta/2$ we find
\be 
C(t)=\frac{1}{2\omega}\frac{\cos (t\omega)}{\sinh (\beta\omega/2)}\;,
\ee
which is the previous $SU(2)$ result for $j=1/2$ and $\alpha=\omega$, up to operator normalization (which should be fixed at initial times). Other $j$ are e.g. achieved by considering a different number of uncoupled harmonic oscillators with the same frequencies.

We also remark that in this class of models, the right assignation of $\alpha$ is temperature independent. This implies that e.g. introducing temperature in a free system does not change the complexity/operator growth. It just changes the correct operator normalization at the initial time, but not the operator wavefunction. This is similar to the computation of Lyapunov exponents at high temperature in SYK \cite{Maldacena:2016hyu} (free regime), where it only depends on the coupling constant, the only scale-dependent parameter of the theory.

Using the operator wavefunction \eqref{WFSU2}, we can now compute the Krylov complexity 
\be
K_\Op=\sum^{2j}_{n=0}n|\varphi_n(t)|^2=2j\sin^2(\alpha t).
\ee
Clearly, there is no exponential growth of complexity in this case.  This fits well with the fact that $b_n$'s are not linear in $n$. More precisely, the complexity grows quadratically $K_\Op\sim 2j\alpha^2t^2$ at early times and reaches its maximum at $t=\pi/(2\alpha)$ given by $K^{max}_\Op=2j$. After that it reduces back to zero at $t=\pi/\alpha$. This is the expected behaviour for a complexity measure. The reason is that $t=\pi/(2\alpha)$ is the furthest point in the complexity geometry, as we are going to show shortly. Passing that point in phase space we begin our trip back to the initial state.

As in the case of SL(2,R), we can better observe these complexity features geometrically, by deriving the information metric associated with \eqref{CSSU2}. This is the spherical metric
\be
ds^2=\frac{2jdzd\bar{z}}{(1+|z|^2)^2}=\frac{j}{2}\left(d\theta^2+\sin^2\theta d\phi^2\right).
\ee
We conclude that this particular non-chaotic operator dynamics is associated with geometry of constant positive curvature
\be
R=\frac{4}{j}.
\ee
As before, for large spin $j$ the curvature decreases. The operator growth again gets mapped to a geodesic in this geometry. 

Last but not least, we can evaluate the volume in this information geometry up to $\theta=2\alpha t$ 
\be
V_t=\int^{2\alpha t}_0d\theta \int^{2\pi}_0d\phi\sqrt{g}=2\pi j\sin^2(\alpha t)=\pi K_\Op\;,
\ee
confirming our proposed relation between the volume in the information geometry and the Krylov complexity.
\subsection{Example III: Heisenberg-Weyl} 
The next example is somewhat in between the previous two. It concerns the Heisenberg-Weyl algebra and its associated standard coherent states. We start with the usual creation ($a^\dagger$) and annihilation ($a$) operators, the identity $1$ and the number operator ($\hat{n}=a^\dagger a$). These operators define the following algebra
 \be
[a,a^\dagger]=1,\qquad [\hat{n},a^\dagger]=a^\dagger,\quad [\hat{n},a]=-a,
\ee
with all other commutators vanishing. The infinite dimensional Hilbert space is expanded in the usual orthonormal basis
\be
\ket{n}=\frac{1}{\sqrt{n!}}(a^\dagger)^n\ket{0},
\ee
on which the ladder operators $a^\dagger$ and $a$ act as
\be
a^\dagger\ket{n}=\sqrt{n+1}\ket{n+1},\qquad a\ket{n}=\sqrt{n}\ket{n-1},
\ee
and $n$ is the eigenvalue of the number operator.

\begin{figure}[t!]
  \centering
  \includegraphics[width=8cm]{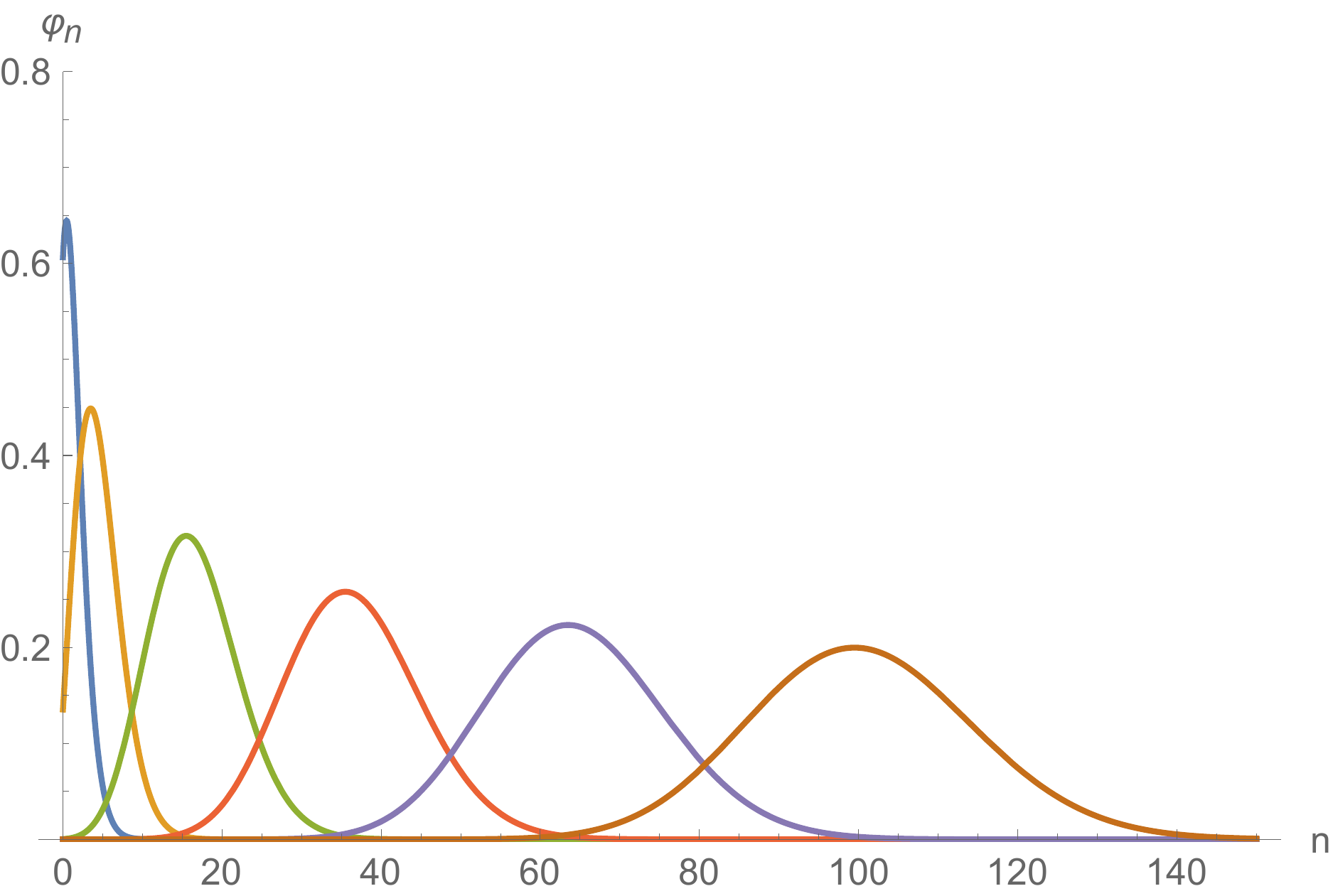}
  \caption{Wavefunctions from the Weyl-Heisenberg coherent states as functions of $n$ for $\alpha t=1,2,4,6,8,10$, from left to right.}
\label{WFnHW}
\end{figure}

Following the general paradigm, these relations allow us to identify the Heisenberg-Weyl Liouvillian, the infinite dimensional Krylov basis and the Lanczos coefficients. These are given by
\be
\mathcal{L}=\alpha(a^\dagger+a),\quad |\Op_n)=\ket{n},\quad b_n=\alpha\sqrt{n}.
\ee
The standard coherent states are defined by the action of the displacement operator on the vacuum state
\be
\ket{z}=D(z)\ket{0},\quad  D(z)=e^{z a^\dagger-\bar{z}a},
\ee
with complex coordinate $z=r e^{i\phi}$. Using the previous algebra one finds
\be
\ket{z}=e^{-|z|^2/2}\sum^\infty_{n=0}\frac{z^n}{\sqrt{n!}}\ket{n}.
\ee
We can now find the operator wavefunction by exploring the relation between the unitary evolution with the Liouvillian and the displacement operator. In particular, by setting $z=i\alpha t$, or $r=\alpha t$ and $\phi=\pi/2$, we write the Heisenberg's operator state in the Krylov space
\be
|\Op(t))=\ket{z=i\alpha t}=e^{i\alpha(a^\dagger+a)t}\ket{0},
\ee 
from where the operator wavefunction is
\be
\varphi_n(t)=e^{-\alpha^2 t^2/2}\frac{\alpha^n t^n}{\sqrt{n!}}, \qquad \sum^\infty_{n=0}|\varphi_n|^2=1.
\ee
It solves the Schrodinger equation \eqref{SchrodingerEq} with the above $b_n$'s and corresponding probabilities form the Poisson distribution. Examples are plotted on Fig. \ref{WFnHW}.\\
Here, the basis is infinite-dimensional but the growth of $b_n$'s is not maximal. Also, the auto-correlation function, in this case, is exponentially decaying
\be
C(t)=\varphi_0(t)=\exp\left(-\alpha^2t^2/2\right).
\ee

With the explicit solution we can compute the Krylov complexity 
\be
K_\Op=\sum^\infty_{n=0}n|\varphi_n(t)|^2=\alpha^2 t^2\;.
\ee
Similarly to SL(2,R) and SU(2), the early time growth is universally proportional to $t^2$ (quadratic), but here it continues as such for all times. 

Finally, the information metric is the flat ($R=0$) complex plane
\be
ds^2_{FS}=dzd\bar{z}=dr^2+r^2d\phi^2.
\ee
The classical trajectory $r=\alpha t$ describing the operator dynamics is again a geodesic in this manifold. Moreover, we  find the same universal relation between the Volume in the information geometry and Krylov complexity
\be
V_t=\int^{\alpha t}_0dr\int^{2\pi}_0d\phi\sqrt{g}=\pi \alpha^2 t^2=\pi K_\Op.
\ee
In the next example, we proceed towards settings more akin to holography. 
\subsection{Example IV: Conformal Field Theories in 2d}
As our last example we consider the case of global symmetry of 2d CFTs \cite{DiF,Fradkin:1996is}. This is an extension of the single $SL (2,R)$  described above and it follows by considering the growth defined by two copies of $SL (2,R)$, corresponding to the global part of the conformal group.  From this perspective, the SYK example is equivalent to a chiral CFT with a single SL(2,R). The Lie algebra is then given by
\bea
\left[ L_0,L_{\pm 1}\right]&=&\mp L_{\pm 1},\qquad \left[ L_1,L_{-1}\right]=2L_0, \nn \\
\left[ \bar{L}_0 ,\bar{L}_{\pm 1}\right]  &=& \mp \bar{L}_{\pm 1},\qquad \left[ \bar{L}_1,\bar{L}_{-1}\right]=2\bar{L}_0\;.
\eea
We will begin with the highest weight state $\ket{h,\bar{h}}$ that is an eigenstate of the CFT Hamiltonian
\be
H\ket{h,\bar{h}}=(L_0+\bar{L}_0) \ket{h,\bar{h}}=(h+\bar{h})\ket{h,\bar{h}},
\ee
This state arises by acting with the mode $\Op_{-h,-\bar{h}}$ of the primary operator $\Op$ with conformal dimension $\Delta=h+\bar{h}$ and spin $s=h-\bar{h}$ on the CFT vacuum
\be
\ket{h,\bar{h}}=\Op_{-h,-\bar{h}}\ket{0,0}.
\ee
Generalization of the coherent states to 2d CFT is now straightforward. We just use two copies of the SL(2,R) displacement operator
\be
\ket{z,h;w,\bar{h}}=D(\xi)\bar{D}(\zeta)\ket{h,\bar{h}},
\ee
where $z$ and $w$ are related to $\xi$ and $\zeta$ respectively as in~(\ref{zxi}). In this case, we need to work with the sum of two Liouvillians (see appendix A) with generally two different coefficients
\be 
\mathcal{L}=\alpha_+\,(L_{-1}+L_{1}),\quad \bar{\mathcal{L}}=\alpha_-\,(\bar{L}_{-1}+\bar{L}_{1})\;.
\ee
For example, in a general (charged) thermal state inner-product with different left and right temperatures $\beta_\pm=T^{-1}_\pm=\beta(1\pm\Omega)$ (e.g. with angular momentum and chemical potential $\Omega$) we may associate $\alpha_\pm=\pi/\beta_\pm$. 

Unitary evolution with the Liouvillian is again a displacement of the initial state. By setting
\bea
z=i\tanh(\alpha_+ t),\qquad w=i\tanh(\alpha_- t)\;,
\eea
this leads to the coherent state or operator wavefunction
\bea
\vert\Op(t))&=&\ket{z=i\tanh(\alpha_+ t),h;w=i\tanh(\alpha_- t),\bar{h}}\nonumber\\
&=&\sum^\infty_{n,m=0}\varphi_{n,m}(t)\ket{h,n;\bar{h},m}\nonumber\;.
\eea
The total wavefunction $\varphi_{n,m}(t)=\varphi_{n}^{\alpha_+}(t)\varphi_{m}^{\alpha_-}(t)$ is a product of the ``left" and ``right" wavefunctions:
\bea\label{WFsCFT}
\varphi_{n}^{\alpha_+}(t)&=&\sqrt{\frac{\Gamma(2h+n)}{n!\Gamma(2h)}}\frac{\tanh^n(\alpha_+ t)}{\cosh^{2h}(\alpha_+ t)}\nonumber\\
\varphi_{m}^{\alpha_-}(t)&=&\sqrt{\frac{\Gamma(2\bar{h}+m)}{m!\Gamma(2\bar{h})}}\frac{\tanh^m(\alpha_- t)}{\cosh^{2\bar{h}}(\alpha_- t)}\;.
\eea
The evolution of $\varphi_{n,m}(t)$ is again described by the Schrodinger equation with two pairs of ``left" and ``right" SL(2,R) Lanczos coefficients. Details are given in appendix A.

This product solution and its $n=m=0$ components lead to a consistent two-point function in CFT
\bea
C(t)&=&(\Op(t)\vert\Op(0))\nonumber\\
&\simeq&\left(\cosh\frac{\pi t}{\beta_+}\right)^{-2h}\left(\cosh\frac{\pi t}{\beta_-}\right)^{-2\bar{h}}.
\eea
Recall that, for holographic CFTs, this two point correlator can be computed in the standard way from gravity by the exponent of the length of a geodesic that stretches between the two sides of the eternal black hole (see e.g. \cite{Hartman:2013qma}).

The Krylov complexity is now the sum of the expectation values of the position in the left and right chains (see also next section)
\be
K_\Op=\sum_{n,m}(n+m)|\varphi_{n,m}(t)|^2.
\ee
In terms of the total conformal dimension $\Delta=h+\bar{h}$ and the spin $s=h-\bar{h}$ this becomes
\bea
K_\Op&=&\Delta\left[\sinh^2\left(\frac{\pi t}{\beta_+}\right)+\sinh^2\left(\frac{\pi t}{\beta_-}\right)\right]\nonumber\\
&+&s\left[\sinh^2\left(\frac{\pi t}{\beta_+}\right)-\sinh^2\left(\frac{\pi t}{\beta_-}\right)\right].
\eea
Clearly, for $\beta_+\neq \beta_-$, Krylov complexity is sensitive to the operator's spin s.

Finally, the information geometry consists of two copies of the Euclidean Poincar\'e disc~(\ref{FSSU11}). The classical trajectory corresponds to a geodesic in this product manifold and the Krylov complexity is proportional to the volume as in previous examples.

Note that this generalization to 2d CFTs seems completely determined by symmetries (see more discussion in  \cite{Dymarsky:2021bjq}). This universality is a simple consequence of the operators that we chose to describe the growth of. For a free CFT, we could have chosen a momentum mode instead, and the Krylov approach would look like the case of SU(2), instead of SL(2,R). We can also imagine considering composite CFT operators and/or consider the thermal CFT on the circle. Such setups will require more detailed information about the CFT spectrum etc, and they will distinguish between chaotic and non-chaotic CFTs. \\
Moreover, in 2d CFTs, we could also study the less universal Liouvillian dynamics based on the Virasoro algebra. Even though we leave this as an interesting future direction, in appendix B we consider a simpler but non-trivial example of SL(2,R) subalgebras of the Virasoro algebra given by $\{L_{-k},L_0,L_k\}$ for some fixed k \cite{Witten:1987ty}. Already in this case we end up with Krylov complexity that  depends on the central charge $c$ of the CFT
\be
K_\Op=2h_k\sinh^2(\alpha_k t),
\ee
where
\be
h_k=\frac{c}{24}\left(k-\frac{1}{k}+\frac{24h}{ck}\right),\qquad \alpha_k=k\alpha,
\ee
and Lanczos coefficients are also asymptotically linear $b_n\simeq \alpha_k n$.\\
We will return to discussion of CFT generalizations at the very end.
\section{V. Complexity algebra and geometry}\label{sec:V}
Previously we have analyzed specific examples related to different groups. In this section, we come back to a more general discussion of the Lanczos coefficients in the light of symmetry. We argue that there exists a natural algebra associated with operator dynamics and Krylov complexity, and that the closure of this algebra on different levels provides another way towards finding potential sets of Lanczos coefficients. In particular, we will again reproduce our previous results from this angle.

The logic proceeds as follows. As described above, the action of the Liouvillian in the Krylov basis yields two terms \eqref{LinKB} and suggests a definition of ``generalized ladder operators'' 
\be
\mathcal{L}=\tilde{L}_++\tilde{L}_-,
\ee
where for simplicity we absorbed $\alpha$ into the ladder operators of the previous section $\tilde{L}_\pm=\alpha L_\pm$ such that
\bea 
 \tilde{L}_+|\Op_n)=b_{n+1}|\Op_{n+1}),\quad  \tilde{L}_-|\Op_n)=b_{n}|\Op_{n-1}).\,~
\eea
The algebra generated by the generalized ladder operators $\tilde{L}_+$ and $\tilde{L}_-$ is simply equivalent to the algebra generated by the Liouvillian and the operator $\mathcal{B}$, defined as
\be 
\mathcal{B}=\tilde{L}_+-\tilde{L}_-\;.
\ee
By definition, action of this anti-Hermitian operator on the Krylov basis is 
\be
\mathcal{B}|\Op_n)=-b_n|\Op_{n-1})+b_{n+1}|\Op_{n+1})\;.\label{AE2}
\ee
We now want to explore the following question. What happens when we start commuting these two operators? From their definitions, we can easily derive the action of the commutator, that we name $\tilde{K}$, in the Krylov basis. Using \eqref{LinKB} and \eqref{AE2} we obtain
\be
\tilde{K}\equiv[\mathcal{L},B]|\Op_n)=2(b^2_{n+1}-b^2_n)|\Op_{n}).
\ee
This operator turns out to be diagonal in the Krylov basis with eigenvalues $\tilde{k}(n)=2(b^2_{n+1}-b^2_n)$.

Given this generic algebraic structure, we now entertain a ``simplicity'' hypothesis. This hypothesis demands that these three operators close an algebra that we may call a ``complexity algebra''. This enforces the following constraint on the commutator eigenvalues
\be
\tilde{k}(n)=An+B\;,
\ee
for some constants $A$ and $B$, implying that $\tilde{k}(n)$ grows at most linearly in $n$. We then conclude that this hypothesis (closure of the algebra) provides a recurrence equation for the Lanczos coefficients
\be
2(b^2_{n+1}-b^2_n)=An+B.
\ee
A general solution to this equation is given by (the positive root)
\be
b_{n}=\sqrt{\frac{1}{4}An(n-1)+\frac{1}{2}Bn+C}\;,
\ee
with $C$ also being an arbitrary constant. Furthermore, requiring $b_0=1$, which holds for any operator growth, fixes this constant to $C=0$. This family of Lanczos coefficients was also derived in \cite{Dymarsky:2019elm} from the Toda chain approach.\\  We see that the hypothesis does not allow the Lanczos coefficients to grow faster than $n$. It would be interesting to see if imposing the closure of the algebra at a later level, by allowing the complexity algebra to include more operators generated by $\mathcal{L}$ and $\mathcal{B}$, still enforces the universal linear bound.

Note that the examples considered earlier all fall within the simplicity hypothesis. For instance, for $SL(2,R)$ we have that
\be
\mathcal{L}=\alpha(L_{-1}+L_1),\quad \mathcal{B}=\alpha(L_{-1}-L_1),\quad \tilde{K}=4\alpha^2L_0,
\ee
and hence the eigenvalue
\be
\tilde{k}_{sl(2,R)}(n)=4\alpha^2(n+h).
\ee
Moreover, we can observe a simple relation between the commutator $\tilde{K}$ and the Krylov complexity operator, namely
\be \label{kkr}
\tilde{K}=4\alpha^2 (\hat{K}_\Op+h)\;.
\ee
They are the same up to a constant and a proportionality factor. In particular, they both grow exponentially with the same growth rate/Lyapunov exponent. This suggests that the operator $L_0$  (or more generally the energy in CFTs) may also be a good candidate for operator complexity or a witness of the operator growth. This proposal was put forward in \cite{Magan2018} and the present results provide a firmer ground for this idea. Nevertheless, the definition of the operator \eqref{KOper} seems more robust, especially from the point of view of generic systems that we can analyze only numerically.
 
The geometric interpretation of these complexity algebra generators is also very elegant. They are just related to the Killing vectors of the information metric \eqref{FSSU11}, that in our coordinates become
\bea
L_0&=&i\partial_\phi,\nn\\
L_{-1}&=&-ie^{-i\phi}\left[\coth(\rho)\partial_\phi+i\partial_\rho\right],\nn\\
L_{1}&=&-ie^{i\phi}\left[\coth(\rho)\partial_\phi-i\partial_\rho\right].
\eea
The operators $(\mathcal{L},\mathcal{B},\tilde{K})$ are built from these generators and satisfy the same algebra. They are therefore associated with the isometries of the information metric. In particular, $\tilde{K}$, almost equal to the Krylov complexity operator, generates translations in $\phi$. Since the difference between $\tilde{K}$ and $K$ is a constant, which just produces non-physical overall phases, we conclude that the Krylov complexity operator is also the generator of translations in $\phi$. In addition, the geometric picture shows that (absolute value of) the expectation value of the operator $\mathcal{B}$ also grows exponentially in the course of operator dynamics (i.e., with $\rho=2\alpha t$ and $\phi=\pi/2$).

In complete analogy, for SU(2) we have
\be
\mathcal{L}=\alpha(J_{+}+J_-),\quad \mathcal{B}=\alpha(J_{+}-J_-),\quad \tilde{K}=-4\alpha^2J_0,
\ee
and the eigenvalues of $\tilde{K}$ becomes
\be
\tilde{k}_{su(2)}(n)=-4\alpha^2(n-j).
\ee
Again this implies a simple relation with the Krylov complexity operator
\be
\tilde{K}=-4\alpha^2 (\hat{K}_\Op-j)\;,
\ee
and both $\tilde{K}$ and the Krylov complexity operator generate rotations in the information metric.

Finally, for the Heisenberg-Weyl algebra the appropriate assignation is
\be
\mathcal{L}=\alpha(a^\dagger+a),\quad \mathcal{B}=\alpha(a^\dagger-a),\quad \tilde{K}=2\alpha^21,
\ee
providing the eigenvalue
\be
\tilde{k}_{HW}(n)=2\alpha^2.
\ee
In this case, the commutator is proportional to the identity. Therefore, the relation to Krylov complexity operator is not just a simple constant shift and appears less natural.

We believe that this new perspective will serve as a solid starting point for a systematic approach to the classification of various operator dynamics. This classification could start by specifying the number of operators we need to add to the Liouviilian and the operator $\mathcal{B}$ to close an algebra. This program may follow by analyzing the possible representations of such a complexity algebra.  Certainly, new examples associated with other Lie groups as well as their deformations will serve as important data points in this direction.
\section{VI. Relation to Geometric Complexity: Particle on a group}\label{sec.VI}
In this section, we provide a bridge between operator dynamics and Krylov complexity, and the geometric approach to computational/circuit complexity. This approach stands out in its similarities to the way physicists think. It was pioneered by Nielsen and collaborators \cite{Nielsen1,Nielsen2,Nielsen3} and, more recently, attracted significant attention with prospective applications to holographic complexity (see e.g. \cite{Jefferson:2017sdb,Chapman:2017rqy,Caputa:2018kdj,Chagnet:2021uvi,Koch:2021tvp,Ge:2019mjt,Balasubramanian:2021mxo,Balasubramanian:2019wgd,Brandao:2019sgy} as well as  \cite{Caputa:2017yrh,Boruch:2021hqs,Belin:2018bpg,Chen:2020nlj} for some of the alternative definitions). The main elegant idea in Nielsen's works is to think about quantum circuits as paths in the manifold of unitary transformations. These paths are determined by different choices of instantaneous quantum gates, characterised by time-dependent Hamiltonians.

On one hand, linking these two ideas is important from a physics point of view, given the recent activity concerning the relation between complexity and black hole physics \cite{SusskindQC,SStanford,Brown1,Brown2,Susskind:2018tei,Magan2018}. On the other hand, this connection can sharpen the operational meaning of the operator wavefunction and the Krylov complexity.

The bridge is built upon the previously described connection to generalized coherent states. 
To make our points clear, we start from the transition amplitude between a coherent state $\ket{z_i}$ at some initial time $t_i$ and a coherent state $\ket{z_f}$ for time $t_f$, defined as
\be
T(z_f,t_f;z_i,t_i)=\bra{z_f}\exp\left(-iH(t_f-t_i)\right)\ket{z_i}.
\ee
We can write a path integral representation of these transition amplitudes, see \cite{Klauder:1960kt}
\be
T(z_f,t_f;z_i,t_i)=\int d\mu[z(t)]e^{iS},
\ee
where $d\mu[z(t)]$ is an appropriate functional measure (an invariant measure on the coset space associated with the problem) and $S$ is an action functional for the paths 
\be
S=\int^{t_f}_{t_i}L(z(t),z'(t),\bar{z}(t),\bar{z}'(t))dt.
\ee
For our discussion on complexity, we only need the classical approximation to this propagator and the classical equations of motion from $S$. In particular, see  \cite{Klauder:1960kt}, the Lagrangian above takes the following standard form
\be
L=\langle z|i\partial_t-H|z\rangle\equiv\langle z|i\partial_t|z\rangle-\mathcal{H}(z,\bar{z}).
\ee
In this formula, the first term is simply $\partial_t=z'\partial_z+\bar{z}'\partial_{\bar{z}}$. It acts on the coherent state before the overlap is computed (see appendix C). In the second term we have defined the ``classical Hamiltonian" $\mathcal{H}$ as
\be
\mathcal{H}(z,\bar{z})=\bra{z}H\ket{z},
\ee
usually referred to as the ``symbol'' of the quantum Hamiltonian $H$. For now, we keep $H$ general, but we will consider particular choices momentarily.

The classical Euler-Lagrange equations of motion for the two variables $z(t)$ and $\bar{z}(t)$ associate with this action are
\be
\frac{d}{dt}\left(\frac{\partial L}{\partial z'}\right)-\frac{\partial L}{\partial z}=0,\quad \frac{d}{dt}\left(\frac{\partial L}{\partial \bar{z}'}\right)-\frac{\partial L}{\partial \bar{z}}=0.
\ee
Equivalently, following the Hamiltonian approach we can write the same equations in terms of Poisson brackets
\be
z'(t)=\{z,\mathcal{H}(z,\bar{z})\},\qquad \bar{z}'(t)=\{\bar{z},\mathcal{H}(z,\bar{z})\}.
\ee
It is now straightforward to check that for our three main examples, with general classical Hamiltonian $\mathcal{H}(z,\bar{z})$, the previous equations of motion can be written in terms of the following Poisson brackets. For the group $SL(2,R)$, corresponding to the hyperbolic phase space geometry, we define
\be
\{A,B\}=i\frac{(1-|z|^2)^2}{2h}\left(\frac{\partial A}{\partial \bar{z}}\frac{\partial B}{\partial z}-\frac{\partial A}{\partial z}\frac{\partial B}{\partial \bar{z}}\right).
\ee
For $SU(2)$, the phase space is the sphere and we have 
\be
\{A,B\}=i\frac{(1+|z|^2)^2}{2j}\left(\frac{\partial A}{\partial \bar{z}}\frac{\partial B}{\partial z}-\frac{\partial A}{\partial z}\frac{\partial B}{\partial \bar{z}}\right).
\ee
Finally, the standard flat one for the Weyl-Heisenberg scenario reads
\be
\{A,B\}=i\left(\frac{\partial A}{\partial \bar{z}}\frac{\partial B}{\partial z}-\frac{\partial A}{\partial z}\frac{\partial B}{\partial \bar{z}}\right).
\ee
Let us now return to the quantum Hamiltonians and their classical counterparts. Since the appropriate group acts transitively in phase space, the Hamiltonian must be an element of the Lie algebra (see \cite{RevModPhys.62.867} for a review of this type of symmetry dynamics). This condition preserves the coherence of the states through evolution. A general Hamiltonian for the SL(2,R) example is then written as
\be
H_{SL(2,R)}=aL_0+bL_1+cL_{-1},
\ee
where $a,b,c$ are arbitrary constants. They can be time-dependent. Indeed, in connections with circuit/Nielsen complexity, these parameters might depend non-trivially on circuit time, defining the instantaneous gate at each time. However, for the discussion in this section, it is sufficient to consider them constant.

We can now compute the expectation value of such a generic operator, ``the symbol" (see App.D) and obtain the following classical Hamiltonian in phase space
\be
\mathcal{H}(z,\bar{z})=\frac{h}{1-|z|^2}\left(a(1+|z|^2)+2bz+2c\bar{z}\right).
\ee
The classical Hamilton equations of motion associated with this Hamiltonian turn out to be the complex Riccati equations and, after we express everything in terms of $\rho$ and $\phi$ coordinates, they become
\bea
\rho'(t)&=&i\left(be^{i\phi(t)}-ce^{-i\phi(t)}\right),\nn\\
\phi'(t)&=&-a-\coth(\rho(t))\left(be^{i\phi(t)}+ce^{-i\phi(t)}\right).\,~
\eea
Our previous solution  $\rho=2\alpha t$ and $\phi=-\pi/2$ describing the operator wavefunction is indeed a solution of these equations with $a=0$ and $c=b=\alpha$. This is consistent with the expected Hamiltonian $H=\alpha(L_{-1}+L_1)$. The minus sign difference in $\phi=\pm \pi/2$ can be traced back to the forward vs backward time evolution with Hamiltonian vs Liouvillian. Dynamics with such Hamiltonians was also discussed in \cite{Tada:2019rls} and this context is may related to our choice of the inner product. We leave more detailed explorations in this direction for future works.

Similarly, for the SU(2) case, the most general coherence-preserving Hamiltonian is written as
\be
H_{SU(2)}=aJ_0+bJ_++cJ_-.
\ee
Taking the expectation value in the coherent state basis we obtain the classical Hamiltonian
\be
\mathcal{H}_{SU(2)}=-ja\cos(\theta)+j\sin(\theta)(be^{-i\phi}+c e^{i\phi}),
\ee
leading to the following equations of motion
\bea
\theta'(t)&=&i\left(ce^{i\phi(t)}-be^{-i\phi(t)}\right),\nn\\
 \phi'(t)&=&-a-(be^{-i\phi(t)}+ce^{i\phi(t)})\cot(\theta(t)).
\eea
Again, the solution described earlier is a solution to these equations corresponding to $a=0$ and Hamiltonian $H=\alpha(J_++J_-)$, as it should be.

Finally, for standard Heisenberg-Weyl symmetry we generically have
\be
H_{HW}=a\hat{n}+b\hat{a}^\dagger+c\hat{a}+d1,
\ee
with arbitrary constants $a,b,c,d$. The classical Hamiltonian is then
\be
\mathcal{H}_{HW}=ar^2+r(be^{-i\phi}+ce^{i\phi})+d,
\ee
and the equations follow
\bea
r'(t)&=&-\frac{i}{2}(be^{-i\phi(t)}-ce^{i\phi(t)}),\nn\\
\phi'(t)&=&-a-\frac{1}{2r}(be^{-i\phi(t)}+ce^{i\phi(t)}).
\eea
Setting $\phi=-\pi/2$ as well as $a=0$ and $b=c=\alpha$ (or $r=\alpha t$) corresponds to motion of a particle on this phase space with Hamiltonian $H_{HW}=\alpha(a^\dagger+a)$.

We conclude that the operator wavefunctions considered earlier, including the example of SYK, can be simply mapped to classical motions i.e., solutions of the classical Hamilton equations of motion in the appropriate generalized coherent state phase space. The classical Hamiltonians above follow directly by taking the expectation value in the generalized coherent states of our proposed form of the Liouvillian $\mathcal{L}=\alpha (L_++L_-)$. This way, we can not only think about operator growth geometrically but also naturally regard unitary Liouvillian evolution as a quantum circuit 
\be
|\Op(t))=e^{i\mathcal{L}t}|\Op).
\ee

With these results in mind, let us return to Nielsen's approach. In this framework, after assigning a particular (highly non-unique) cost function to the instantaneous gates, one can estimate the computational complexity of the task by finding the length of the minimal geodesic in the geometry of unitaries. The ambiguity in the cost functions somewhat parallels the freedom in choosing the inner product to turn the operator algebra into a Hilbert space. However, the geodesic length is not obviously related to the operator complexity. Naively, one is tempted to identify the information geometry (Fubini-Study metric) with Nielsen's metric, but as we saw before the geodesic length between the origin and $\rho=2\alpha t$ (at fixed $\phi=\pi/2$) grows only linearly in time. Indeed, as we saw before, it is the phase space Volume in the Fubini-Study metric that Krylov complexity measures. 

Still, one can interpret Krylov complexity in terms of a geodesic length. This fact comes from the universal relation between the $\mathcal{F}_1$ norm and Krylov complexity  (see \eqref{F1N} in Appendix D). Indeed, for phase-space displacements in the angular direction we have
\be
\mathcal{F}_1=|\langle z|\delta z\rangle|=K_\Op d\phi,
\ee
This can be interpreted as the Nielsen complexity, defined with $\mathcal{F}_1$ cost functions \cite{Magan2018,Caputa:2018kdj,Bueno:2019ajd,Erdmenger:2020sup}, of the circuit that takes us from trajectory $(\rho=2\alpha t,\phi=\pi/2)$ to a nearby geodesic with $(\rho=2\alpha t,\phi=\pi/2+\delta\phi)$. This in turn is very closely related to the definition of classical chaos that we discussed in the introduction.
\section{VII. Quantum Information tools for operator growth} 
This last section is devoted to contrasting the evolution of Krylov complexity with more conventional quantum information tools. For this, we step again on the connection between operator dynamics and coherent states. More concretely, a certain two-mode representation of the displacement operator will allow us to derive a density matrix associated with the evolving operator. Then, instead of quantifying complexity with expectation values of operators, such as the Krylov complexity, we will explore it with different quantum information tools. As new outcomes, we will discuss traces of the operator growth in entanglement measures, define a notion of operator proper temperature that connects to the physics of black holes and to quantum optics. This last outcome will suggest a way to contrast these theoretical problems with experiments.

Below we will concentrate on the ``chaotic" example of SL(2,R), and focus on the time dependence of the different quantities, comparing their growth with Krylov complexity. Most of these results hold for other coherent states as well and here we only survey the most important findings. Further details with non-chaotic examples will be described in \cite{Patramanis}.

We start by representing the SL(2,R) generators in terms of two oscillator modes as
\be
L_{-1}=a^\dagger_1a^\dagger_2,\quad L_1=a_1a_2,\quad L_0=\frac{1}{2}(a^\dagger_1 a_1+a^\dagger_2a_2+1),
\ee
where the creation $a^\dagger_i$ and annihilation  $a_i$ operators satisfy the Weyl-Heisenberg algebra. In this representation, the displacement operator $D(\xi)$ becomes the standard two-mode squeezing operator frequently used in theoretical as well as experimental quantum optics, see e.g. \cite{Agarwal2012QuantumOptics}. Then we consider the so-called k-photon added/subtracted states 
\be
\ket{z,k}\equiv\mathcal{N}(a_2)^kS(\xi)\ket{0,0}\;,
\ee
in which there is a difference of k-excitations between the two modes. Using the standard Bogoliubov transformation we can expand this state in the two oscillator Fock space basis as
\be
\ket{z,k}=(1-|z|^2)^{\frac{k+1}{2}}\sum^{\infty}_{n=0}z^n \sqrt{\frac{\Gamma(k+1+n)}{n!\Gamma(k+1)}}\ket{n+k,n}.
\ee
In the amplitudes of this state we recognise those of the SL(2,R) coherent states. One just needs to perform the identification $k+1=\eta=2h$, while the phase space coordinates z's remain unchanged. In this squezeed representation, the coherent states are entangled states. Also, in this form, the Krylov basis  is the standard two-oscillator Fock space
\be
|\Op_n)=\ket{n+k,n}=\frac{(a^\dagger_1)^{n+k}}{\sqrt{(n+k)!}}\frac{(a^\dagger_2)^n}{\sqrt{n!}}\ket{0,0}.
\ee
In this two-mode representation, we can think about the operator wavefunction $|\Op(t))$ as a ``perturbed'' thermofield double state. Tracing out the second oscillator we arrive at the following density matrix 
\be
\rho^{(k)}_1=Tr_2\left(\ket{z,k}\langle z,k|\right)=\sum^\infty_{n=0} \lambda_n\ketbra{n+k}.
\ee
Its eigenvalues are precisely the probabilities in the Krylov basis
\be
\lambda_n=|\varphi_n(t)|^2=\frac{\Gamma(2h+n)}{n!\Gamma(2h)}(1-|z|^2)^{2h} |z|^{2n}, 
\ee
where we remind that in order to describe operator growth we need to assign $z=i\tanh(\alpha t)$. 

This description of the operator growth process allows us to assign a ``proper temperature'' to the operator. To this end, and for simplicity, we analyze the special case of $k=0$ (or $h=1/2$). Then defining
\be
e^{-\beta(t)\omega}=\tanh^2(\alpha t),
\ee
the mixed state $\rho^{(k)}_1$ is just the thermal state of the harmonic oscillator with inverse temperature $\beta(t)$. This is a temperature $T(t)=1/\beta(t)$ naturally associated with operator growth. At large times we find that this operator temperature behaves as
\be 
T(t)\xrightarrow[\alpha t\gg 1]{} \frac{e^{2\alpha t}}{4}\;,
\ee
growing exponentially fast with the right Lyapunov exponent. Quite interestingly, this is the expected behaviour of proper temperatures/energies of infalling perturbations into a black hole, which is just universally controlled by the near-horizon redshift, determined by the time-time component of the black hole metric.

This density matrix representation of the operator growth also allows us to explore it using tools from quantum information. First of all, the K-entropy, defined in \cite{Barbon:2019wsy}, is just the standard von-Neumann entropy of $\rho^{(k)}_1$, now appearing as an entanglement entropy 
\be
S_\Op=-\sum_n|\varphi_n|^2\log(|\varphi_n|^2),
\ee
between the two modes of the squeezed state. The analytic answer is obtained, for example, by setting $k=0$ and and it grows linearly at late times
\be
S_{\Op}=2\log\left(\cosh(\rho)\right)-2\log\left(\tanh(\rho)\right)\sinh^2(\rho)\simeq 2\alpha t,
\ee 
with a proportionality factor equal to the Lyapunov exponent. This behaviour is reminiscent of that of a Kolmogorov-Sinai entropy, whose rate of growth is upper bounded by the sum of Lyapunov exponents.

More generically we can compute Renyi entropies
\be
S^{(q)}_\Op=\frac{1}{1-q}\log\left(\sum_n|\varphi_n|^{2q}\right),
\ee
and for $k=0$ these are given by
\be
S^{(q)}_\Op=\frac{1}{q-1}\log\left(\cosh^{2q}(\alpha t)(1-\tanh^{2q}(\alpha t))\right).\label{Renyk0}
\ee

One more interesting quantity that has been studied recently in various contexts \cite{DeBoer:2018kvc,Kawabata:2021hac,Nandy:2021hmk} is the capacity of entanglement
\be
\mathcal{C}_{\Op}=\lim_{q\to 1}q^2\partial^2_q\left[(1-q)S^{(q)}_\Op\right].
\ee
In particular, in \cite{Nandy:2021hmk} it was shown to be a useful new probe of local operators.
For $k=0$, we can easily compute it analytically from \eqref{Renyk0} and it becomes
\be
\mathcal{C}_{\Op}=\sinh^2(2\alpha t)(\log(\tanh(\alpha t)))^2.
\ee
This way, capacity of entanglement also grows for early times  $\alpha t\simeq 0$ as
\be
\mathcal{C}_{\Op}\simeq 4\alpha^2t^2\log(\alpha t)^2,
\ee
but then saturates to $1$ exponentially fast as $t\to\infty$ (with twice the Lyapunov exponent)
\be
\mathcal{C}_{\Op}\xrightarrow[\alpha t\gg 1]{}1-\frac{4}{3}e^{-4\alpha t}.
\ee
The physical interpretation of this saturation (``thermalisation") is not yet clear to us but it indicates that not all the probes must necessarily grow/decay (linearly or exponentially) in order to extract certain universal features from their evolution.

Related to entanglement entropy, a useful measure of ``quantumness" of a state considered in quantum optics is entanglement negativity \cite{Vidal:2002zz}. For operator growth, it can be written in terms of the operator wavefunctions as 
\be
E_\mathcal{N}(\rho)=2\log\left(\sum_n|\varphi_n|\right).
\ee
In the simplest case with $k=0$, it becomes precisely
\be
E_\mathcal{N}(\rho)=2\alpha t,
\ee
resembling the growth of a geodesic length from the origin of the hyperbolic disc.

Finally, when quantifying complexity of quantum states it is natural to expect that different distance measures also play a significant role (see e.g. \cite{Miyaji:2015woj,DiGiulio:2020hlz,Yang:2018nda}). For example, relative entropy between $\sigma=\ket{k}\langle k|$ and $\rho^{(k)}_1$ is given by
\bea
S(\sigma|\rho^{(k)}_1)&=&Tr(\sigma\log(\sigma))-\Tr(\sigma\log(\rho^{(k)}_1))\nonumber\\
&=&-2\log(|\varphi_0(t)|)=4h\log(\cosh(\alpha t)),\nn\\
\eea
growing linearly with time for late times. Another example is the fidelity
\bea
F(\sigma,\rho)&=&Tr\left(\rho^{1/2}\sigma \rho^{1/2}\right)=|\varphi_0(t)|\nonumber\\
&=&\cosh^{-2h}(\alpha t)\simeq 2^{2h}e^{-2h\alpha t},
\eea
that again decays with time. Interestingly, both of these measures are simply expressed by $\varphi_0(t)$.

The main message from this section is that these QI tools can be directly written in terms of $\varphi_n(t)$'s and are sensitive to the Lyapunov exponent $2\alpha$. The example of proper temperatures and their relation to black hole physics is quite intriguing. In addition, from the geometric perspective, entanglement negativity or the late time relative entropy are bounded by the geodesic length from the origin up to the radius $\rho(t)$. More discussion and detailed analysis of the quantum information tools for the operator growth will be presented in \cite{Patramanis}.
\section{VIII. Discussion: CFT Generalizations}
After analyzing our examples, it is clear that most of the discussion has been constrained by the universal kinematics of the underlying symmetries. This was already discussed in \cite{Dymarsky:2021bjq}. Of course, for tests and applications to genuine holography in two and higher dimensions, we will need to focus on less universal setups/operators that are sensitive to more detailed aspects of the spectrum and dynamics.

Some generalizations were already put forward in the original work \cite{Parker:2018yvk}. These come under the name of q-complexities. In fact, the authors argued that Krylov complexity bounds the growth of the OTOC, that arise by choosing inner products of the type $([V,\Op(t)]|[V,\Op(t)])$, for arbitrary operators $\Op(t)$ and $V$. Another interesting generalization was studied in \cite{Kar:2021nbm} and focused on a certain fixed energy bandwidth. Below, we discuss more of such new directions that are natural from the perspective of CFTs and holography.

Firstly, extending the symmetry discussion in 2d CFTs to the full Virasoro algebra as well as the conformal algebra in higher dimensions would be very interesting. This would parallel the circuit complexity setup proposed in \cite{Caputa:2018kdj}, see also \cite{Erdmenger:2020sup,Chagnet:2021uvi}. For instance, investigating the Lanczos coefficients and their scaling with $n$ for the full Virasoro group, as well as $W_n$ or BMS, would teach us important lessons about holographic CFTs. We hope to report on these developments in future works. 

A more immediate generalization arises as follows. We can consider analyzing the growth, not of a single operator $\mathcal{O}(t)$, but a product of operators $\mathcal{O}_1(t)\mathcal{O}_2(t)$ with generally different conformal dimensions. This problem requires more detailed non-universal information about the CFT. For instance, the auto-correlation function is now a four-point function
\bea
C(t)&=&(\mathcal{O}_1(t)\mathcal{O}_2(t)\vert \mathcal{O}_1(0)\mathcal{O}_2(0))\nonumber\\&\equiv &\langle e^{ H\beta/2}\mathcal{O}_1(t)\mathcal{O}_2(t)^\dagger e^{-H\beta/2}\mathcal{O}_1(0)\mathcal{O}_2(0) \rangle_\beta,\nn\\
\eea
which opens a path to relate Krylov complexity and the more conventional approach to quantum chaos based on the OTOC correlators in QFTs. To study the growth of $\mathcal{O}_1(t)\mathcal{O}_2(t)$ we may resort to the Operator Product Expansion (OPE) in CFT. Two CFT operators can be fused and written as a sum of local operators. For our purposes we need to take the limit of coincident points in the OPE. Given that the coefficient of a certain operator $\mathcal{O}$ of dimension $\Delta$ in the OPE of $\mathcal{O}_1(x)\mathcal{O}_2(y)$ is down by powers of $\vert x-y\vert^{2\Delta}$ as $x\rightarrow y$, the fusion is controlled by the operator $\mathcal{O}(t)$ with lowest dimension appearing in the OPE. In this limit we can write
\be 
\mathcal{O}_1(t')\mathcal{O}_2(t)\underset{t'\rightarrow t} \longrightarrow F(t-t')\,\mathcal{O}(t).\;
\ee
The proportionality factor $F(t-t')$ will go away once we normalize the initial state. We conclude that the growth of the composite operator $\mathcal{O}_1(t)\mathcal{O}_2(t)$ is described by the growth of the single operator $\mathcal{O}'$ with the lowest $\Delta$ appearing in the OPE. To analyze it, we can use the results of the previous section.

Another natural generalization emerges by analyzing the Krylov complexity of $\mathcal{O}_1(0)\mathcal{O}_2(t)$, as time evolves, i.e., by considering composite operators at non-coincident locations. In this setup, the OPE does not collapse to a single operator but is expanded in terms of OPE blocks, see \cite{Fradkin:1996is,BCzech:2016xec,deBoer:2016pqk}. This suggests considering first the Krylov complexity of OPE blocks. In the light of the present article, the block Liouvillian is expected to take the generic form~(\ref{liug}) and the block complexity may be related to volumes in kinematic space \cite{BCzech:2016xec,deBoer:2016pqk}. We leave these interesting problems for future work as well.

Last but not least, we may be interested in building the Krylov basis and Krylov complexity for the time evolution of more general initial states (not necessarily simple operators). These include time evolution of boundary and TFD states of CFTs, evolution of states dual to local operators in the bulk \cite{Goto:2017olq}, or microscopic SYK high energy states such as the ones considered in \cite{perm}. In the first two examples, we naturally encounter objects related to the return amplitudes \cite{Cardy:2014rqa}
\be
\mathcal{F}(t)=(\Psi|e^{i\mathcal{L}t}|\Psi),\,
\ee
that in turn are closely related to spectral form factors (see e.g. \cite{Dyer:2016pou}) used in the studies of chaos and operator growth. Using them as inputs (generalizing the auto-correlation function) of the Lanczos algorithm and studying their Krylov complexity is definitely one the most important future directions.
\section{IX. Conclusions} 
In the context of operator and circuit complexity, and their applications to theoretical physics, the are two problems that stand out. The first, and in our opinion probably the most important, is to frame the context in a way that transparently connects to the ``physicist perspective''. The second is to have a formulation that allows for progress on a technical level. 

The objective of this article has been to develop a unifying approach to these and other related problems. Concerning the complexity of operators we advocated for the Krylov complexity and the Lanczos approach as a natural starting point. Concerning the physics, it has been crucial to analyze systems from the perspective of their symmetry, generalized coherent states and their associated geometries. These two, a priori distinct, fields share a common ground when one notices that the Liouvillian/Hamiltonian of the system can be written in the Lanczos basis as the sum of ladder operators (and more generally diagonal symmetry generators). Dynamics of the system with that type of ``symmetry''  is typically equivalent to the classical evolution of a ``particle'' in the appropriate phase space.

This equivalence allows us to make progress in various directions. It provides a more transparent, physical meaning to Krylov complexity operator. It turns out to be related to the vector fields generating isometries of the classical phase space geometry.
Equivalently, it is an element of the Lie algebra of the symmetry group of the system. Moreover, its expectation value is the volume defined by the classical motion on the geometry. From a more technical perspective, the present identification simplifies enormously the computation of Lanczos coefficients. 

Finally, we want to end with several open questions. First, there is an interesting consequence of the relation between operator complexity and classical evolution in appropriate phase spaces. In several cases of interest the problem can be stated in a language that makes contact with the field of quantum optics. One might enjoy the thought that the classical process that faithfully represents the operator growth wavefunction in the SYK model could be analyzed experimentally using photons. This promising avenue deserves more development and we hope to come back to it soon.

From a more theoretical perspective, the relation between Krylov complexity and phase space volumes calls for a deeper understanding. In parallel with generalized coherent states \cite{coherent2}, it will be interesting to consider operator dynamics governed by Liouvillians that are arbitrary combinations of the algebra elements and explore the relation with volume (operator) further. The role of the volume certainly resonates with the holographic complexity proposal made in \cite{SStanford}, and we expect it will teach us new insights about black hole interiors. 

Finally, the relation between operator growth and classical motion described above, together with notions of quantum chaos and Lyapunov exponents based on the out-of-time correlators may indeed allow for a more physical derivation of the bound on quantum chaos \cite{Maldacena:2015waa,Murthy:2019fgs}.

\medskip
{\bf Acknowledgements.} We wish to thank Jan Boruch, Diptarka Das, Felix Haehl, Sinong Liu, Wolfgang Muck, Pratik Nandy, Milosz Panfil, Krzysztof Wohlfeld, Onkar Parrikar, Tadashi Takayanagi, Claire Zukowski for discussions and comments and especially, Anatoly Dymarsky, Dongsheng Ge and Joan Simon for discussions and comments that advanced this work. The work of P.C and D.P is supported by NAWA “Polish Returns 2019” and NCN Sonata Bis 9 grants. The work of J.M is supported by a DOE QuantISED grantDE-SC0020360 and the Simons Foundation It From Qubit collaboration (385592).

\appendix
\section{Appendix A: Lanczos Algorithm}\label{appx:Setup}
In this first appendix, we carry on explicitly the Lanczos algorithm with Liouvillian built from the elements of the Lie algebra. Instead of recurring from the intuition of coherent states, we will proceed carefully through each step of the iterative process.

The algorithm starts with the initial operator and the second element of the Lanczos basis
\be
|\Op_0):=|\Op),\qquad |\Op_1):=b^{-1}_1\mathcal{L}|\Op_0),
\ee
where $b_1=(\Op_0\mathcal{L}|\mathcal{L}\Op_0)^{1/2}$. The algorithm then constructs an orthogonal basis of states iteratively as
\be
|A_n)=\mathcal{L}|\Op_{n-1})-b_{n-1}|\Op_{n-2}),
\ee 
which are then normalized
\be
|\Op_n)=b^{-1}_n|A_n),\qquad b_n=(A_n|A_n)^{1/2}.
\ee
In addition to the basis states $|\Op_n)$, this algorithm outputs the so-called Lanczos coefficients $b_n$.

For simplicity, we will analyze in detail the example of the SL(2,R) algebra \eqref{SL2R} with associated Liouvillian \eqref{LLSU2R}. In the first step of the Lanczos algorithm we choose an initial operator/state 
\be
|\Op_0)=|\Op)=\ket{h}.
\ee
This is the highest weight state
\be
L_0\ket{h}=h\ket{h},\qquad L_{1}\ket{h}=0.
\ee
We find it conveneint to think about this example as a ``chiral CFT" where 
\be
\ket{h}=\Op_{-h}\ket{0},
\ee
where $\ket{0}$ is the vacuum and $\Op_{-h}$ is a mode of a primary operator with (chiral) dimension $h$. These primary field modes satisfiy the following commutation relations with the generators of the conformal algebra
\be
[L_m,\Op_{n}]=((h-1)m-n)\Op_{n+m}.
\ee
In particular
\be
[L_m,\Op_{-h}]=(h(m+1)-m)\Op_{-h+m},\label{CROn}
\ee
where for us $m\in\{-1,0,1\}$. These field modes also satisfy their own algebra, generally written for two quasi-primary operators $\Op^{(i)}$ and $\Op^{(j)}$ as
\bea
[\Op^{(i)}_{m},\Op^{(j)}_n]&=&\delta_{m+n,0}d^{ij}\binom{m+h_i-1}{2h_i-1}\nn\\
&+&\sum_{k}C^{ij}_k p^{ij}_k(m,n)\Op^{(k)}_{m+n},\label{CROPS}
\eea
where  $d^{ij}$ are the two-point function coefficients (that are usually set to $\delta_{ij}$ by normalizing the operators), $C^{ij}_k$ are the three-point function coefficients and $p^{ij}_k(m,n)$ are some universal polynomials specified by the CFT data (see e.g. \cite{DiF}).

The next step is to normalize the second basis state of the Lanczos basis, namely
\be
|\Op_1)=b^{-1}_1\mathcal{L}|\Op_0)=b^{-1}_1\alpha L_{-1}\ket{h}.
\ee
The normalization, i.e. the first Lanczos coefficient, is computed from the commutation relation of the SL(2,R) algebra. It is given by
\be
b_1=(\Op_0\mathcal{L}|\mathcal{L}\Op_0)^{1/2}=\alpha\sqrt{\langle h|L_{1}L_{-1}|h\rangle}=\alpha\sqrt{2h}.
\ee
This way we have
\be
|\Op_1)=\frac{1}{\sqrt{2h}}L_{-1}\ket{h}\equiv\ket{h,1}.
\ee
Notice that in this chiral CFT we could equivalently interpret this second basis state in terms of the commutator of the mode $\Op_{-h}$ with the following ``Hamiltonian''
\be
\mathcal{L}\Op_{-h}=[H,\Op_{-h}],\qquad H=\alpha(L_{-1}+L_1).
\ee
Indeed, from the commutation relations \eqref{CROn} it is clear that
\be
[H,\Op_{-h}]=\alpha\left(\Op_{-(h+1)}+(2h-1)\Op_{-h+1}\right),
\ee
hence
\be
|\mathcal{L}\Op_0)=[H,\Op_{-h}]\ket{0}=\alpha \Op_{-(h+1)}\ket{0}.
\ee
The normalization of this state is obtained from \eqref{CROPS} (with $d^{ij}=\delta_{ij}$)
\be
b_1=\alpha\langle 0|[\Op_{h+1},\Op_{-(h+1)}]|0\rangle^{1/2}=\alpha\sqrt{2h}.
\ee
We now move to the third step and find
\bea
|A_{2})&=&\mathcal{L}|\Op_{1})-b_{1}|\Op_{0}),\nn\\
&=&\frac{\alpha}{\sqrt{2h}}\left(L^2_{-1}\ket{h}+L_1L_{-1}\ket{h}\right)-\alpha\sqrt{2h}\ket{h},\nn\\
&=&\frac{\alpha}{\sqrt{2h}}L^2_{-1}\ket{h},
\eea
where in the second equality we computed the product of $L_1L_{-1}$ using the SL(2,R) commutator. This canceled the last term. Computing the Lanczos coefficient
\be
b_2=(A_2|A_2)^{1/2}=\frac{\alpha}{\sqrt{2h}}\langle h|L^2_{1}L^2_{-1}|h\rangle^{1/2},
\ee
we obtain
\be
b_2=\alpha\sqrt{2(2h+1)},
\ee
where we have used the general result
\be
\langle h|L^n_{1}L^n_{-1}|h\rangle=n!\frac{\Gamma(2h+n)}{\Gamma(2h)}.
\ee
We then arrive at the third basis vector
\be
|\Op_2)=b^{-1}_2|A_2)=\frac{1}{\sqrt{4h(2h+1)}}L^2_{-1}\ket{h}\equiv \ket{h,2}.
\ee
Continuing this procedure we can derive the full $SL(2,R)$ result for the Lanczos coefficients
\be
b_n=\alpha\sqrt{n(2h+n-1)},
\ee
as well as the Krylov basis vectors
\be
|\Op_n)=\ket{h,n}=\sqrt{\frac{\Gamma(2h)}{n!\Gamma(2h+n)}}L^{n}_{-1}\ket{h},
\ee
These vectors are orthonormal $(\Op_m|\Op_n)=\delta_{n,m}$.

It is also useful to repeat the Lanczos Algorithm for the two-mode representation of the SL(2,R) Liouvillian that we used in the main text
\be
\mathcal{L}=\alpha\left(a^\dagger_2a^\dagger_1+a_2a_1\right).
\ee
This operator can be written in terms of two oscillator modes with standard commutation relations 
\be
[a_i,a^\dagger_i]=1,\qquad [a_i,a^\dagger_{j}]=0\quad \text{for}\quad i\neq j,
\ee
with $i,j=1,2$. In what follows, the two-mode vacuum state is anihilated by both operators $a_i$
\be
a_i\ket{0,0}=0,
\ee
and we introduce the n-particle orthonormal states
\be
\ket{n}_i=\frac{(a^\dagger_i)^n}{\sqrt{n!}}\ket{0}_i.
\ee
Without causing confusion, we will drop the subscript from the states, with the understanding that operators act on their appropriate Hilbert spaces. We then have
\be
a_i\ket{n}=\sqrt{n}\ket{n-1},\qquad a^\dagger_i\ket{n}=\sqrt{n+1}\ket{n+1}.
\ee
In this version, the Lanczos algorithm starts from a super-operator
\be
\Op^{(k)}=\frac{(a^\dagger_1)^k}{\sqrt{k!}},
\ee
and the corresponding two-mode state
\be
|\Op^{(k)}_0)=\Op^{(k)}\ket{0,0}=\ket{k,0}.
\ee
In quantum optics, this state is usually called the k-photon added state since the difference between the excitation numbers in each mode is $k$. We then proceed with the algorithm and apply the Liouvillian to define the second basis vector
\be
|O^{(k)}_1)=b^{-1}_1\mathcal{L}|O^{(k)}_0)=\frac{\alpha\sqrt{k+1}}{b_1}\ket{k+1,1},
\ee
where
\be
\ket{k+1,1}=a^\dagger_1a^\dagger_2\ket{k,0}.
\ee
The normalization (the first Lanczos coefficient) is now given by
\be
b_1=\sqrt{(O^{(k)}_0\mathcal{L}|\mathcal{L}O^{(k)}_0)}=\alpha\sqrt{k+1},
\ee
so that
\be
|O^{(k)}_1)=\ket{k+1,1}.
\ee
In the next step  we write
\bea
|A_2)&=&\mathcal{L}\ket{k+1,1}-b_1\ket{k,0},\nonumber\\
&=&\alpha\sqrt{2(k+2)}\ket{k+2,2}+\alpha\sqrt{k+1}\ket{k,0}-b_1\ket{k,0}\nonumber\\
&=&\alpha\sqrt{2(k+2)}\ket{k+2,2},
\eea
where
\be
\ket{k+2,2}=\frac{(a^\dagger_1)^2}{\sqrt{2!}}\frac{(a^\dagger_2)^2}{\sqrt{2!}}\ket{k,0}.
\ee
The normalization fixes the second Lanczos coefficient to
\be
b_2=(A_2|A_2)^{1/2}=\alpha\sqrt{2(k+2)},
\ee
and finally we get the third basis vector
\be
|\Op^{(k)}_2)=\ket{k+2,2}.
\ee
This way, repeating the above procedure, we construct an orthonormal Krylov basis
\be
|\Op^{(k)}_n)=\ket{k+n,n}=\frac{(a^\dagger_1)^{n}}{\sqrt{n!}}\frac{(a^\dagger_2)^{n}}{\sqrt{n!}}\ket{k,0},
\ee
with Lanczos coefficients
\be
b_n=\alpha\sqrt{n(k+n)}.\label{bn}
\ee
The relation to the previous representation arises by the simple assignation $2h=k+1$.

Analogous computations can be done for general 2d CFTs. In this scenario the symmetry group is $SU(2,R)_L\otimes SU(2,R)_R$ and we have two Liouvillians; left (chiral) $\mathcal{L}$ and right (anti-chiral) $\bar{\mathcal{L}}$. The starting point now is the highest weight state
\be
|\Op^{h,\bar{h}})=\Op_{-h,-\bar{h}}\ket{0}_L\otimes\ket{0}_R
\ee
and we are interested in computing the wavefunction in the Krylov basis that describes the evolution of the operator
\bea
|\Op^{(h,\bar{h})}(t))&=&e^{i(\mathcal{L}+\bar{\mathcal{L}})t}|\Op^{h,\bar{h}}),\nn\\
&=&\sum^\infty_{n,m=0}i^{n+m}\varphi_{n,m}(t)|\Op^{(h,\bar{h})}_{n,m}).
\eea
The Lanczos algorithm can be followed exactly as before, but with two sets of Lanczos coefficients defined from the actions of the Liouvillians in the Krylov basis
\bea
\mathcal{L}|\Op^{(h,\bar{h})}_{n,m})&=&b_{n}|\Op^{(h,\bar{h})}_{n-1,m})+b_{n+1}|\Op^{(h,\bar{h})}_{n+1,m}),\nn\\
\bar{\mathcal{L}}|\Op^{(h,\bar{h})}_{n,m})&=&\bar{b}_{m}|\Op^{(h,\bar{h})}_{n,m-1})+\bar{b}_{m+1}|\Op^{(h,\bar{h})}_{n,m+1}).
\eea
The Schrodinger equation becomes
\bea
\partial_t\varphi_{n,m}(t)&=&b_n\varphi_{n-1,m}(t)-b_{n+1}\varphi_{n+1,m}(t)\nn\\
&+&\bar{b}_m\varphi_{n,m-1}(t)-\bar{b}_{m+1}\varphi_{n,m+1}(t).
\eea
We can check that for
\be
b_n=\alpha\sqrt{n(n+2h-1)},\qquad \bar{b}_m=\bar{\alpha}\sqrt{m(m+2\bar{h}-1)},
\ee
this equation is solved by the product of the wavefunctions \eqref{WFsCFT}. \\
Let us finally point that we can start the Lanczos algorithm from a more general initial state where the Liouvillian has a non-zero expectation value i.e., diagonal part.  Equivalently, one may be interested in constructing the Krylov basis for a general Hamiltonian evolution \cite{Lanczosbook}. In such cases, the algorithm can be repeated similarly with
\be
|A_{n+1})=(\mathcal{L}-a_{n})|\Op_n)-b_n|\Op_{n-1}),\quad |\Op_n)=b^{-1}_n|A_n),
\ee
and the two sets of coefficients are defined as
\be
a_n=(\Op_n|\mathcal{L}|\Op_n),\qquad b_n=(A_n|A_n)^{1/2}.
\ee
From these expressions we can easily determine the action of the Liouvillian in the Krylov basis 
\be
\mathcal{L}|\Op_n)=a_n|\Op_n)+b_{n+1}|\Op_{n+1})+b_n|\Op_{n-1}),
\ee
and derive a discrete Schrodinger equation for the wavefunctions
\be
\partial_t\phi_n(t)=i\left(a_n\phi_n+b_n\phi_{n-1}+b_{n+1}\phi_{n+1}\right),
\ee
that define the general state
\be
|\Op(t))=\sum_{n}\phi_n|\Op_n).
\ee
In symmetry setups such as SL(2,R), $a_n$'s can e.g. be naturally associated with eigenvalues of $L_0$.
\section{Appendix B: $SL(2,R)$ subalgebras of Virasoro}
In this appendix we want to consider another example of our symmetry proposal. This is a natural variant of the SL(2,R) algebra scenario. Namely, instead of the global part of the conformal group, let us now consider the Virasoro algebra
\be\label{Virasoro}
[L_n,L_{m}]=(n-m)L_{n+m}+\frac{c}{12}n(n^2-1)\delta_{n+m,0},
\ee
and focus on a different subset of three generators, namely $L_0$ and two Virasoro modes 
$L_k$ and $L_{-k}$ for a fixed integer $k>0$. Using~(\ref{Virasoro}), we see these three generators form a closed algebra since
\be
[L_k,L_{-k}] = 2kL_0+\frac{c}{12}k(k^2-1),
\ee
as well as
\be
[L_0,L_{\pm k}] = \mp k L_{\pm k}.
\ee
By redefining
\be
\tilde{L}_0=\frac{1}{k}\left(L_0+\frac{c}{24}(k^2-1)\right),\qquad \tilde{L}_{\pm 1}=\frac{1}{k}L_{\pm k},
\ee
this is the standard $SL(2,R)$ algebra \eqref{SL2R}.

Given this observation, we can consider the following displacement operator
\be
D_k(\xi)=e^{\xi L_{-k}-\bar{\xi}L_{k}}.
\ee
Using the BCH formula and $\xi=re^{i\phi}$ this can be written in factorized form
\bea
D_k(\xi)&=&e^{e^{i\phi}\frac{\tanh(k r)}{k}L_{-k}}e^{-\frac{2}{k}\log(\cosh(kr))\left(L_0+\frac{c}{24}(k^2-1)\right)}\nonumber\\
&\times&e^{-e^{-i\phi}\frac{\tanh(k r)}{k}L_{k}}.
\eea
Applying it to the eigenstate $\ket{h}$ satisfying
\be
L_0\ket{h}=h\ket{h},\qquad L_{k}\ket{h}=0,
\ee
we derive
\be\label{cohk}
D_k(\xi)\ket{h}=\frac{1}{\cosh^{2h_k}(kr)}\sum^{\infty}_{n=0}e^{in\phi}\frac{\tanh^n(kr)}{n!k^n}L^{n}_{-k}\ket{h}.
\ee
where
\be
h_k=\frac{1}{k}\left(h+\frac{c}{24}(k^2-1)\right).
\ee
Using the following result
\be
\bra{h}L^n_kL^n_{-k}\ket{h}=n!k^{2n}\frac{\Gamma(2h_k+n)}{\Gamma(2h_k)},
\ee
we can introduce the orthonormal basis
\be
\ket{h,nk}\equiv \sqrt{\frac{\Gamma(2h_k)}{n!k^{2n}\Gamma(2h_k+n)}}L^{n}_{-k}\ket{h},
\ee
such that
\be
\bra{h,nk}h,mk\rangle=\delta_{n,m}.
\ee
We can now write the coherent state~(\ref{cohk}) in this basis and obtain
\be
\ket{z,h,k}=\sum^{\infty}_{n=0}e^{in\phi}\frac{\tanh^n(kr)}{\cosh^{2h_k}(kr)} \sqrt{\frac{\Gamma(2h_k+n)}{n!\Gamma(2h_k)}}\ket{h,nk}.
\ee
To connect with the Lanczos approach we first notice that
\bea
L_0\ket{h,nk}&=&(h+nk)\ket{h,nk},\nn\\
L_{-k}\ket{h,nk}&=&k\sqrt{(n+1)(2h_k+n)}\ket{h,(n+1)k},\nn\\
L_{k}\ket{h,nk}&=&k\sqrt{n(2h_k+n-1)}\ket{h,(n-1)k}.
\eea
This structure now allows us to define the Liouvillian and Krylov basis
\be
\mathcal{L}_k=\alpha(L_{-k}+L_{k}),\qquad |\Op_n)=\ket{h,nk},\label{LiuvK}
\ee
together with the Lanczos coefficients
\be
b_n=k\alpha\sqrt{n(2h_k+n-1)}.
\ee
Let us point out that Hamiltonians of the type \eqref{LiuvK} also appeared recently in \cite{Hu:2020suv} but the more precise link remains to be understood.\\
These coefficients satisfy the algebraic relation
\be
b^2_{n+1}-b^2_n=2k^2\alpha^2(h_k+n).
\ee
For large $n$ they grow as
\be
b_n\simeq k\alpha n+k\alpha\frac{2h_k-1}{2}+O(1/n).
\ee
Consequently, we can compute the Krylov complexity in this example
\be
K_\Op=\sum_{n}n|\varphi_n(t)|^2=2h_k\sinh^2(k\alpha t),
\ee
and conclude that the growth rate is now generalized to $\alpha_k=k\alpha$. Interestingly, this result is sensitive to the central charge c of the CFT. More physical aspects of complexity and entanglement in these states will be considered in \cite{DongshengPC}.
\section{Appendix C: CFT Correlators}\label{appx:}
In this appendix, we review/collect the basic material about thermal two-point correlators in 2d CFTs that appeared in the main text.

In Euclidean CFTs in 2d, the basic objects of interest are correlation functions of quasi-primary operators. These operator transform under chiral $z\to f(z)$ and anti-chiral conformal transformations  $\bar{z}\to \bar{f}(\bar{z})$ as
\be
\Op (z,\bar{z})\to (f'(z))^h(\bar{f}'(\bar{z}))^{\bar{h}}\Op(f(z),\bar{f}(\bar{z})),\label{OPTR}
\ee
where the coefficients $h$ and $\bar{h}$ are related to the conformal dimension $\Delta$ and the spin $s$ of the operator by
\be
\Delta=h+\bar{h},\qquad s=h-\bar{h}.
\ee
From these finite transformations we can infer their infinitesimal counterparts. Under $z\to f(z)\simeq z+\epsilon(z)$ (and similarly for $\bar{z}$) we have
\be
\delta_{\epsilon,\bar{\epsilon}}\Op(z,\bar{z})=(h\partial_z\epsilon+\epsilon\partial_z+\bar{h}\partial_{\bar{z}}\bar{\epsilon}+\bar{\epsilon}\partial_{\bar{z}})\Op(z,\bar{z}).
\ee 
The two-point functions of the quasi-primary operators $\Op_i$ and $\Op_j$ in the CFT vacuum state (this state is the state annihilated by the Virasoro modes $L_k\ket{0}=0$ for $k\ge -1$)
\be
G(z_1,z_2)=\langle  0|\Op_i (z_1,\bar{z}_1)\Op_j (z_2,\bar{z}_2)|0\rangle,
\ee
 are fully determined by global conformal symmetry. Indeed, the invariance under inifinitesimal transformations with $\epsilon(z)=z^{k+1}$ for $k=-1,0,1$ forces them to satisfy the following differential equations
 \bea
\sum_i\partial_{z_i}G=\sum_{i}(z_i\partial_{z_i}+h_i)G=\sum_{i}(z^2_i\partial_{z_i}+2h_i z_i)G=0,\nn\\
 \eea
and a similar set for the anti-chiral variables $\bar{z}_i$. These equations can be solved by
\be
G(z_1,z_2)=\frac{d_{ij}\delta_{h_{i},h_j}\delta_{\bar{h}_i,\bar{h}_j}}{z^{2h_i}_{12}\bar{z}^{2\bar{h}_i}_{12}},
\ee
with $z_{ij}=z_i-z_j$. In most of examples one normalizes the operators appropriately such that $d_{ij}=\delta_{ij}$.

CFT's two-point functions on the Euclidean cylinder can now be obtained by the exponential conformal map, together with the transformation rule of the quasi-primary operators \eqref{OPTR}. This procedure allows us to obtain a universal two-point function for a CFT on a circle (finite size). Similarly, if we take the time of the cylinder to be compact with period $\beta$, we can use the same trick and arrive at the universal correlator for a CFT on a line at finite temperature.

More concretely, considering the two exponential conformal maps
\be
f(z)=\exp\left(\frac{2\pi}{\beta_+}z\right),\qquad \bar{f}(\bar{z})=\exp\left(\frac{2\pi}{\beta_-}\bar{z}\right),
\ee
the two-point correlator in a CFT on a line at chiral temperature $T_+=1/\beta_+$ and anti-chiral $T_-=1/\beta_-$ becomes
\bea
\langle\Op(z_1,\bar{z}_1)\Op(z_2,\bar{z}_2)\rangle&=&\left(\frac{\beta_+}{\pi}\sinh\left(\frac{\pi z_{12}}{\beta_+}\right)\right)^{-2h}\nn\\
&\times&\left(\frac{\beta_-}{\pi}\sinh\left(\frac{\pi \bar{z}_{12}}{\beta_-}\right)\right)^{-2\bar{h}}.\label{CFTC2B}
\eea
All these results concern Euclidean CFTs. However, when studying operator growth we employed the result for the Lorentzian two-point function. This Lorentzian continuation appeared in the survival amplitude or autocorrelation function of the Lanczos approach
\bea
(\Op(t)|\Op)&\equiv &\langle e^{\frac{\beta}{2} H}\Op(t) e^{-\frac{\beta}{2} H}\Op\rangle_\beta,\nn\\
&=&\langle e^{\frac{\beta}{2} H}e^{iHt}\Op e^{-iHt} e^{-\frac{\beta}{2} H}\Op\rangle_\beta,
\eea
where in the second step we used the definition of the Heisenberg operator. The reason why this correlation function comes about was described in detail in section II.

This Lorentzian correlator can be obtained by analytic continuation of the Euclidean one. More concretely we just write
\be
(\Op(t)|\Op)=\langle e^{\left(\frac{\beta}{2}+it\right) H}\Op e^{-\left(\frac{\beta}{2}+it\right) H}\Op\rangle_\beta.
\ee
This is just the previous euclidean correlator \eqref{CFTC2B}, after the identifications $z=x+i\tau$, $\bar{z}=x-i\tau$ and substitution $\tau\rightarrow\beta/2+it$, so that
\bea
z_{12}&=&x_{12}+i\left(\frac{\beta_+}{2}+it\right),\qquad \bar{z}_{12}=x_{12}-i\left(\frac{\beta_-}{2}+it\right).\nn\\
\eea
\section{Appendix D: Coherent States: Formulas}\label{Appx:C}
In this last appendix, we collect several formulas from generalized coherent states used in the main text. There are many excellent books on this subject. We refer the reader to e.g. \cite{gazeau2009,coherent2}.

We used three main families of coherent states, based on algebra SU(1,1) (or SL(2,R)) i.e., the Perelomov coherent states
\be
\ket{z,h}=(1-|z|^2)^h\sum^\infty_{n=0}z^n\sqrt{\frac{\Gamma(2h+n)}{n!\Gamma(2h)}}\ket{h,n},
\ee
the spin coherent states based on SU(2) algebra
\be
\ket{z,j}=(1+z\bar{z})^{-j}\sum^{2j}_{n=0}z^n\sqrt{\frac{\Gamma(2j+1)}{n!\Gamma(2j-n+1)}}\ket{j,-j+n},
\ee
and the standard coherent states based on the Heisenberg-Weyl algebra
\be
\ket{z}=e^{-|z|^2/2}\sum^\infty_{n=0}\frac{z^n}{\sqrt{n!}}\ket{n}.
\ee
Coherent states form an over-complete basis and are not orthogonal. Namely, in three examples above, for different values of the complex variables we have
\bea
\langle z_2,h|z_1,h\rangle&=&\frac{(1-|z_1|^2)^h(1-|z_2|^2)^h}{(1-z_1\bar{z}_2)^{2h}},\\
\langle z_2,j|z_1,j\rangle&=&\frac{(1+|z_1|^2)^{-j}(1+|z_2|)^{-j}}{(1+z_1\bar{z}_2)^{-2j}},\\
\langle z_2|z_1\rangle&=&\exp\left(-\frac{|z_1|^2+|z_2|^2}{2}+z_1\bar{z}_2\right).
\eea
In the part where the classical motion in phase space was studied and the Lagrangian, was derived, we used 
\be
\langle z,h|i\partial_t|z,h\rangle=i\frac{h(\bar{z}z'-z\bar{z}')}{1-|z|^2},
\ee
for SU(1,1), then
\be
\langle z,j|i\partial_t|z,j\rangle=i\frac{j(\bar{z}z'-z\bar{z}')}{1+|z|^2},
\ee
for SU(2), and finally
\be
\langle z|i\partial_t|z\rangle=\frac{i}{2}\left(\bar{z}z'-z\bar{z}'\right),
\ee
for Weyl-Heisenberg coherent states. Moreover, for the computation of symbols of Hamiltonians, we used the expectation values of individual generators of the Lie algebras in the coherent states, which for SL(2,R) take the form
\bea
\bra{z,h}L_0\ket{z,h}&=&h\frac{1+|z|^2}{1-|z|^2},\nn\\
\bra{z,h}L_{-1}\ket{z,h}&=&\frac{2h\bar{z}}{1-|z|^2},\nn\\
\bra{z,h}L_{1}\ket{z,h}&=&\frac{2hz}{1-|z|^2}.
\eea
From the above relations we determine the expectation values of the complexity algebra generators in these coherent states 
\bea
\bra{z,h}\tilde{K}_\Op\ket{z,h}&=&4h\alpha^2 \cosh(\rho),\nn\\
\bra{z,h}\mathcal{L}\ket{z,h}&=&2h\alpha\cos(\phi)\sinh(\rho),\nn\\
\bra{z,h}\mathcal{B}\ket{z,h}&=&-2ih\alpha\sin(\phi)\sinh(\rho).
\eea
On our trajectory, $\rho=2\alpha t$ and $\phi=\pi/2$ we see that both, $\tilde{K}_\Op$ and the anti-Hermitian $\mathcal{B}$ grow exponentially while the Liouvillian vanishes.\\

Next, the Euler-Lagrange equations where written in terms of Poisson bracket that for $(\rho,\phi)$ coordinates of SU(1,1) became  
\be
\{A,B\}=\frac{1}{h\sinh(\rho)}\left(\frac{\partial A}{\partial \rho}\frac{\partial B}{\partial \phi}-\frac{\partial A}{\partial \phi}\frac{\partial B}{\partial \rho}\right).
\ee
For SU(2) we similarly have
\bea
\bra{z,j}J_0\ket{z,j}&=&-j\frac{1-|z|^2}{1+|z|^2},\nn\\
\bra{z,j}J_+\ket{z,j}&=&j\frac{2\bar{z}}{1+|z|^2},\nn\\
\bra{z,j}J_-\ket{z,j}&=&j\frac{2z}{1+|z|^2}.
\eea
Then, in the $(\theta,\phi)$ coordinates the SU(2) Poisson bracket reads
\be
\{A,B\}=\frac{1}{j\sin(\theta)}\left(\frac{\partial A}{\partial \theta}\frac{\partial B}{\partial \phi}-\frac{\partial A}{\partial \phi}\frac{\partial B}{\partial \theta}\right).
\ee
For Heisenberg-Weyl Lagrangian and Hamiltonian we used
\be
\bra{z}a\ket{z}=a,\quad\bra{z}a^\dagger\ket{z}=\bar{z},\quad \bra{z}\hat{n}\ket{z}=|z|^2.
\ee
Then, in the $r,\phi$ coordinates of the plane, the Weyl-Heisenberg Poisson bracket is
\be
\{A,B\}=\frac{1}{2r}\left(\frac{\partial A}{\partial r}\frac{\partial B}{\partial \phi}-\frac{\partial A}{\partial \phi}\frac{\partial B}{\partial r}\right).
\ee
Next, the symbols for SL(2,R) are related to embedding coordinates of the hyperbolic space as follows
\bea
\bra{z,h}L_0\ket{z,h}&=&\sqrt{2h}\,X_0,\nn\\
\bra{z,h}L_{-1}\ket{z,h}&=&\sqrt{2h}\,X_+,\nn\\
\bra{z,h}L_{1}\ket{z,h}&=&\sqrt{2h}\,X_-,
\eea
and satisfy
\be
-X^2_0+X_+X_-=-h/2.
\ee
We can further parametrize them in terms of $\rho$ and $\phi$ coordinates
\bea
\{X_0,X_+,X_-\}=\sqrt{\frac{h}{2}}\{\cosh(\rho),e^{-i\phi}\sinh(\rho),e^{i\phi}\sinh(\rho)\},\nonumber
\eea
so that the induced metric becomes
\be
ds^2=\frac{h}{2}\left(d\rho^2+\sinh^2\rho d\phi^2\right).
\ee
Similar relations hold for the other two symmetry groups.\\
Last but not least, we discussed the role played by $\mathcal{F}_1$ norms that are given by
\bea\label{F1N}
\langle z,h|\delta z,h\rangle&=&i2h\sinh^2\left(\frac{\rho}{2}\right)d\phi,\\
\langle z,j|\delta z,j\rangle&=&i2j\sin^2\left(\frac{\theta}{2}\right)d\phi,\\
\langle z|\delta z\rangle&=&ir^2d\phi,
\eea 
in SL(2,R), SU(2) and Heisenberg-Weyl examples respectively. For our trajectories $\rho=2\alpha t$, $\theta=2\alpha t$ and $r=\alpha t$, they all have the universal form
\be
\langle z|\delta z\rangle=iK_\Op d\phi.
\ee



\begin{thebibliography}{99}
\bibitem{Lorenz}
E. U. Lorenz, Predictability: Does the Flap of a Butterfly’s Wings in Brazil Set off a Tornado in Texas? American Association for the Advancement of Science, 1979\\ 
 Lorenz, Edward N. (March 1963). "Deterministic Nonperiodic Flow". Journal of the Atmospheric Sciences. 20 (2): 130–141
 
\bibitem{Haake}
Haake F., Gnutzmann S., Kuś M. (2018) Classical Hamiltonian Chaos. In: Quantum Signatures of Chaos. Springer Series in Synergetics. Springer, Cham. 

\bibitem{Berry1989}
M. Berry, "Quantum chaology, not quantum chaos", Physica Scripta 40, 3, 335-336, 1989

\bibitem{Larkin}
A. I. Larkin and Y. N. Ovchinnikov,
``Quasiclassical method in the theory of superconductivity,''
JETP \textbf{28}, 6 (1969)

 \bibitem{Maldacena:1997re}
  J.~M.~Maldacena,
  ``The large N limit of superconformal field theories and supergravity,''
  Adv.\ Theor.\ Math.\ Phys.\  {\bf 2}, 231 (1998)
  [Int.\ J.\ Theor.\ Phys.\  {\bf 38}, 1113 (1999)]

\bibitem{kitaev}
  A.~Kitaev,
  {\em A simple model of quantum holography},
  Talks at KITP, April 7, 2015 and May 27, 2015.

   \bibitem{Shenker:2013pqa}
S.~H.~Shenker and D.~Stanford,
``Black holes and the butterfly effect,''
JHEP \textbf{03} (2014), 067

\bibitem{Shenker:2013yza}
S.~H.~Shenker and D.~Stanford,
``Multiple Shocks,''
JHEP \textbf{12} (2014), 046

\bibitem{Maldacena:2015waa}
J.~Maldacena, S.~H.~Shenker and D.~Stanford,
``A bound on chaos,''
JHEP \textbf{08} (2016), 106
[arXiv:1503.01409 [hep-th]].

\bibitem{Murthy:2019fgs}
C.~Murthy and M.~Srednicki,
``Bounds on chaos from the eigenstate thermalization hypothesis,''
Phys. Rev. Lett. \textbf{123} (2019) no.23, 230606
[arXiv:1906.10808 [cond-mat.stat-mech]].

  \bibitem{susskind}
  Y.~Sekino and L.~Susskind,
  {\em Fast Scramblers},
  JHEP {\bf 0810} (2008) 065,
  [arXiv:0808.2096 [hep-th]]. 

\bibitem{Lashkari:2011yi}
N.~Lashkari, D.~Stanford, M.~Hastings, T.~Osborne and P.~Hayden,
``Towards the Fast Scrambling Conjecture,''
JHEP \textbf{04} (2013), 022

 \bibitem{Barbon1}
  J.~.L.~F.~Barbon and J.~M.~Magan,
  {\em Chaotic Fast Scrambling At Black Holes},
  Phys.\ Rev.\ D.\  {\bf 84} (2011) 106012,
  [arXiv:1105.2581 [hep-th]].
  
  \bibitem{Barbon3}
  J.~.L.~F.~Barbon and J.~M.~Magan,
  {\em Fast Scramblers, Horizons and Expander Graphs},
  JHEP {\bf 1208} (2012) 016,
  [arXiv:1204.6435 [hep-th]].
  
     \bibitem{freeblack}
   J.~M.~Magan,
{\em Black holes as random particles: entanglement evolution in infinite range and matrix models},
 JHEP {\bf 1608} (2016) 081,
 [arXiv:1601.04663 [hep-th]].
 
  \bibitem{sachdev}
   S.~Sachdev and J.~Ye,
   {\em Gapless spin fluid ground state in a random, quantum Heisenberg ferromagnet},
   Phys. \ Rev.\ Lett.\ {\bf 70} (1993) 3339,
   arXiv:cond-mat/9212030 [cond-mat]. 


\bibitem{Roberts:2018mnp}
D.~A.~Roberts, D.~Stanford and A.~Streicher,
``Operator growth in the SYK model,''
JHEP \textbf{06} (2018), 122
[arXiv:1802.02633 [hep-th]].

\bibitem{Qi:2018bje}
X.~L.~Qi and A.~Streicher,
``Quantum Epidemiology: Operator Growth, Thermal Effects, and SYK,''
JHEP \textbf{08} (2019), 012
[arXiv:1810.11958 [hep-th]].

\bibitem{Kudler-Flam:2020yml}
J.~Kudler-Flam, M.~Nozaki, S.~Ryu and M.~T.~Tan,
``Entanglement of Local Operators and the Butterfly Effect,''
[arXiv:2005.14243 [hep-th]].

\bibitem{MacCormack:2020auw}
I.~MacCormack, M.~T.~Tan, J.~Kudler-Flam and S.~Ryu,
``Operator and entanglement growth in non-thermalizing systems: many-body localization and the random singlet phase,''
[arXiv:2001.08222 [cond-mat.str-el]].

\bibitem{Kudler-Flam:2019wtv}
J.~Kudler-Flam, M.~Nozaki, S.~Ryu and M.~T.~Tan,
``Quantum vs. classical information: operator negativity as a probe of scrambling,''
JHEP \textbf{01} (2020), 031

\bibitem{Magan:2020iac}
J.~M.~Mag\'an and J.~Simon,
``On operator growth and emergent Poincar\'e symmetries,''
JHEP \textbf{05} (2020), 071
[arXiv:2002.03865 [hep-th]].

\bibitem{Parker:2018yvk}
D.~E.~Parker, X.~Cao, A.~Avdoshkin, T.~Scaffidi and E.~Altman,
``A Universal Operator Growth Hypothesis,''
Phys. Rev. X \textbf{9} (2019) no.4, 041017

\bibitem{Lanczosbook}
The Recursion Method: Application to Many-Body Dynamics, vol. 23 Science  Business Media
V. Viswanath, G. Müller\\
C.~Lanczos,
``An iteration method for the solution of the eigenvalue problem of linear differential and integral operators,''
J. Res. Natl. Bur. Stand. B \textbf{45} (1950), 255-282

\bibitem{Rabinovici:2020ryf}
E.~Rabinovici, A.~S\'anchez-Garrido, R.~Shir and J.~Sonner,
``Operator complexity: a journey to the edge of Krylov space,''
[arXiv:2009.01862 [hep-th]].

\bibitem{Barbon:2019wsy}
J.~L.~F.~Barb\'on, E.~Rabinovici, R.~Shir and R.~Sinha,
``On The Evolution Of Operator Complexity Beyond Scrambling,''
JHEP \textbf{10} (2019), 264

\bibitem{Jian:2020qpp}
S.~K.~Jian, B.~Swingle and Z.~Y.~Xian,
``Complexity growth of operators in the SYK model and in JT gravity,''
JHEP \textbf{03} (2021), 014

\bibitem{Yin:2020oze}
C.~Yin and A.~Lucas,
``Quantum operator growth bounds for kicked tops and semiclassical spin chains,''
Phys. Rev. A \textbf{103} (2021) no.4, 042414
[arXiv:2010.06592 [cond-mat.str-el]].


\bibitem{Dymarsky:2019elm}
A.~Dymarsky and A.~Gorsky,
``Quantum chaos as delocalization in Krylov space,''
Phys. Rev. B \textbf{102} (2020) no.8, 085137
[arXiv:1912.12227 [cond-mat.stat-mech]].

\bibitem{Dymarsky:2021bjq}
A.~Dymarsky and M.~Smolkin,
``Krylov complexity in conformal field theory,''
[arXiv:2104.09514 [hep-th]].

\bibitem{Carrega:2020jrk}
M.~Carrega, J.~Kim and D.~Rosa,
``Unveiling Operator Growth Using Spin Correlation Functions,''
Entropy \textbf{23} (2021) no.5, 587

\bibitem{Kar:2021nbm}
A.~Kar, L.~Lamprou, M.~Rozali and J.~Sully,
``Random Matrix Theory for Complexity Growth and Black Hole Interiors,''
[arXiv:2106.02046 [hep-th]].


\bibitem{vonKeyserlingk:2017dyr}
C.~von Keyserlingk, T.~Rakovszky, F.~Pollmann and S.~Sondhi,
``Operator hydrodynamics, OTOCs, and entanglement growth in systems without conservation laws,''
Phys. Rev. X \textbf{8} (2018) no.2, 021013

\bibitem{Nahum:2017yvy}
A.~Nahum, S.~Vijay and J.~Haah,
``Operator Spreading in Random Unitary Circuits,''
Phys. Rev. X \textbf{8} (2018) no.2, 021014


\bibitem{Ryu:2006bv}
S.~Ryu and T.~Takayanagi,
``Holographic derivation of entanglement entropy from AdS/CFT,''
Phys. Rev. Lett. \textbf{96} (2006), 181602 

  \bibitem{SusskindQC}
L.~Susskind,
  {\em Computational Complexity and Black Hole Horizons},
  Fortsch. Phys. {\bf 64}, 44-48 (2016),
  [arXiv:1402.5674 [hep-th]].
  
  \bibitem{SStanford}
D.~Stanford and L.~Susskind,
  {\em Complexity and Shock Wave Geometries},
  Phys.~Rev.~D {\bf 90}, 12 (2014),
  [arXiv:1406.2678 [hep-th]].
  
   \bibitem{Brown1}
A.~Brown, D.~Roberts, L.~Susskind, B~Swingle, Y.~Zhao,
  {\em Holographic Complexity Equals Bulk Action?},
 Phys.~Rev.~Lett {\bf 116}, 19 (2016),
[arXiv:1509.07876 [hep-th]].

\bibitem{Brown2}
A.~Brown, D.~Roberts, L.~Susskind, B~Swingle, Y.~Zhao,
  {\em Complexity, action, and black holes},
 Phys.~Rev.~D {\bf 93}, 19 (2016),
[arXiv:1512.04993 [hep-th]].

\bibitem{Aaronson}
S.~Aaronson,
  {\em The Complexity of Quantum States and Transformations: From Quantum Money to Black Holes},
  [arXiv:1607.05256 [hep-th]].

\bibitem{Rangamani:2016dms}
M.~Rangamani and T.~Takayanagi,
``Holographic Entanglement Entropy,''
Lect. Notes Phys. \textbf{931} (2017), pp.1-246

\bibitem{Chen:2021lnq}
B.~Chen, B.~Czech and Z.~z.~Wang,
``Quantum Information in Holographic Duality,''
[arXiv:2108.09188 [hep-th]].

\bibitem{Susskind:2018tei}
Susskind, Leonard,
``Why do Things Fall?,''
[arXiv:11802.01198 [hep-th]].

\bibitem{Zhao:2017iul}
Y.~Zhao,
``Complexity and Boost Symmetry,''
Phys. Rev. D \textbf{98} (2018) no.8, 086011

\bibitem{Lin:2019qwu}
H.~W.~Lin, J.~Maldacena and Y.~Zhao,
``Symmetries Near the Horizon,''
JHEP \textbf{08} (2019), 049

\bibitem{Barbon:2020uux}
J.~L.~F.~Barbon, J.~Martin-Garcia and M.~Sasieta,
``A Generalized Momentum/Complexity Correspondence,''
JHEP \textbf{04} (2021), 250

   \bibitem{Magan2018}
J.~Magan,
  {\em Black holes, Complexity and quantum chaos},
 JHEP \textbf{09} (2018), 043
arXiv:1805.05839 [hep-th].

\bibitem{Haehl:2020zgs}
F.~M.~Haehl and Y.~Zhao,
``Size and momentum of an infalling particle in the black hole interior,''
JHEP \textbf{21} (2020), 056

\bibitem{Iliesiu:2021ari}
L.~V.~Iliesiu, M.~Mezei and G.~S\'arosi,
``The volume of the black hole interior at late times,''
[arXiv:2107.06286 [hep-th]].

\bibitem{DeBoer:2019kdj}
J.~De Boer and L.~Lamprou,
``Holographic Order from Modular Chaos,''
JHEP \textbf{06} (2020), 024

\bibitem{RevModPhys.62.867}
Zhang, Wei-Min and Feng, Da Hsuan and Gilmore, Robert
Coherent states: Theory and some applications,
  Rev. Mod. Phys. 62, 4, 867--927, 1990
  
     \bibitem{coherent1}
 J.~ R.~Klauder and E.~ C.~ G.~ Sudarshan,
 Fundamentals of quantum optics, (Benjamin, New York) 1968. 
  
 \bibitem{coherent2}
A.~M.~Perelomov,
 Commun.\ Math.\ Phys.\ {\bf 26} (1972) 222.  
  
  \bibitem{gazeau2009}
 J.P. Gazeau,  "Coherent States in Quantum Physics", 2009, Wiley

\bibitem{Brown:2016wib}
A.~R.~Brown, L.~Susskind and Y.~Zhao,
``Quantum Complexity and Negative Curvature,''
Phys. Rev. D \textbf{95} (2017) no.4, 045010
[arXiv:1608.02612 [hep-th]].

\bibitem{Auzzi:2020idm}
R.~Auzzi, S.~Baiguera, G.~B.~De Luca, A.~Legramandi, G.~Nardelli and N.~Zenoni,
``Geometry of quantum complexity,''
Phys. Rev. D \textbf{103} (2021) no.10, 106021

\bibitem{Flory:2020dja}
M.~Flory and M.~P.~Heller,
``Conformal field theory complexity from Euler-Arnold equations,''
JHEP \textbf{12} (2020), 091

\bibitem{Basteiro:2021ene}
P.~Basteiro, J.~Erdmenger, P.~Fries, F.~Goth, I.~Matthaiakakis and R.~Meyer,
``Quantum Complexity as Hydrodynamics,''
[arXiv:2109.01152 [hep-th]].

 \bibitem{Lin:2019kpf}
H.~W.~Lin and L.~Susskind,
``Complexity Geometry and Schwarzian Dynamics,''
JHEP \textbf{01} (2020), 087
[arXiv:1911.02603 [hep-th]].

  \bibitem{Barbon2}
  J.~.L.~F.~Barbon and J.~M.~Magan,
  {\em Fast Scramblers Of Small Size},
  JHEP {\bf 1110} (2011) 035,
  [arXiv:1106.4786 [hep-th]].
  
  \bibitem{Barbon4}
   J.~.L.~F.~Barbon and J.~M.~Magan,
  {\em Fast Scramblers And Ultrametric Black Hole Horizons},
  JHEP {\bf 11} (2013) 163,
  [arXiv:1306.3873 [hep-th]].


\bibitem{PLaine_2016}
Laine, Mikko and Vuorinen, Aleksi
``Basics of Thermal Field Theory,''
Lecture Notes in Physics, Springer International Publishing, 2016


\bibitem{Maldacena:2016hyu}
J.~Maldacena and D.~Stanford,
``Remarks on the Sachdev-Ye-Kitaev model,''
Phys. Rev. D \textbf{94} (2016) no.10, 106002

   \bibitem{DiF}
P. Di Francesco, P. Mathieu, D. Senechal (1997)
"Conformal field theory"

 \bibitem{Fradkin:1996is}
   Fradkin, E. S. and Palchik, M. Ya.,
  {\em Conformal quantum field theory in D-dimensions},
Mathematics and Its applications {\bf 376} (1996), Springer


\bibitem{Hartman:2013qma}
T.~Hartman and J.~Maldacena,
``Time Evolution of Entanglement Entropy from Black Hole Interiors,''
JHEP \textbf{05} (2013), 014

\bibitem{Witten:1987ty}
E.~Witten,
``Coadjoint Orbits of the Virasoro Group,''
Commun. Math. Phys. \textbf{114} (1988), 1


 \bibitem{Nielsen1}
  M.~A.~Nielsen,
{\em A geometric approach to quantum lower bounds},
 [arXiv:0502070 [quant-ph]].
  
 \bibitem{Nielsen2}
   M.~A.~Nielsen, M.~R.~Dowling, M.~Gu and A.~C.~Doherty
{\em Quantum computation as geometry},
Science {\bf 311}, 1133 (2006),
 [arXiv:0603161 [quant-ph]].
 
  \bibitem{Nielsen3}
 M.~R.~Dowling and M.~A.~Nielsen,
{\em The geometry of quantum computation},
 [arXiv:0701004 [quant-ph]].


\bibitem{Jefferson:2017sdb}
R.~Jefferson and R.~C.~Myers,
``Circuit complexity in quantum field theory,''
JHEP \textbf{10} (2017), 107
[arXiv:1707.08570 [hep-th]].

\bibitem{Chapman:2017rqy}
S.~Chapman, M.~P.~Heller, H.~Marrochio and F.~Pastawski,
``Toward a Definition of Complexity for Quantum Field Theory States,''
Phys. Rev. Lett. \textbf{120} (2018) no.12, 121602
[arXiv:1707.08582 [hep-th]].

  \bibitem{Caputa:2018kdj}
P.~Caputa and J.~M.~Magan,
``Quantum Computation as Gravity,''
Phys. Rev. Lett. \textbf{122} (2019) no.23, 231302
[arXiv:1807.04422 [hep-th]].

\bibitem{Chagnet:2021uvi}
N.~Chagnet, S.~Chapman, J.~de Boer and C.~Zukowski,
``Complexity for Conformal Field Theories in General Dimensions,''
[arXiv:2103.06920 [hep-th]].

\bibitem{Koch:2021tvp}
R.~d.~Koch, M.~Kim and H.~J.~R.~Van Zyl,
``Complexity from Spinning Primaries,''
[arXiv:2108.10669 [hep-th]].

\bibitem{Ge:2019mjt}
D.~Ge and G.~Policastro,
``Circuit Complexity and 2D Bosonisation,''
JHEP \textbf{10} (2019), 276
  
 \bibitem{Balasubramanian:2021mxo}
V.~Balasubramanian, M.~Decross, A.~Kar, Y.~Li and O.~Parrikar,
``Complexity growth in integrable and chaotic models,''
JHEP \textbf{07} (2021), 011

\bibitem{Balasubramanian:2019wgd}
V.~Balasubramanian, M.~Decross, A.~Kar and O.~Parrikar,
``Quantum Complexity of Time Evolution with Chaotic Hamiltonians,''
JHEP \textbf{01} (2020), 134
[arXiv:1905.05765 [hep-th]].

\bibitem{Brandao:2019sgy}
F.~G.~S.~L.~Brand\~ao, W.~Chemissany, N.~Hunter-Jones, R.~Kueng and J.~Preskill,
``Models of Quantum Complexity Growth,''
PRX Quantum \textbf{2} (2021) no.3, 030316

\bibitem{Caputa:2017yrh}
P.~Caputa, N.~Kundu, M.~Miyaji, T.~Takayanagi and K.~Watanabe,
``Liouville Action as Path-Integral Complexity: From Continuous Tensor Networks to AdS/CFT,''
JHEP \textbf{11} (2017), 097; 
P.~Caputa, N.~Kundu, M.~Miyaji, T.~Takayanagi and K.~Watanabe,
``Anti-de Sitter Space from Optimization of Path Integrals in Conformal Field Theories,''
Phys. Rev. Lett. \textbf{119} (2017) no.7, 071602

\bibitem{Belin:2018bpg}
A.~Belin, A.~Lewkowycz and G.~S\'arosi,
``Complexity and the bulk volume, a new York time story,''
JHEP \textbf{03} (2019), 044    

\bibitem{Boruch:2021hqs}
J.~Boruch, P.~Caputa, D.~Ge and T.~Takayanagi,
``Holographic path-integral optimization,''
JHEP \textbf{07} (2021), 016; 
J.~Boruch, P.~Caputa and T.~Takayanagi,
``Path-Integral Optimization from Hartle-Hawking Wave Function,''
Phys. Rev. D \textbf{103} (2021) no.4, 046017
 
 \bibitem{Chen:2020nlj}
B.~Chen, B.~Czech and Z.~z.~Wang,
``Query complexity and cutoff dependence of the CFT2 ground state,''
Phys. Rev. D \textbf{103} (2021) no.2, 026015  
   
    
\bibitem{Klauder:1960kt}
J.~R.~Klauder,
``The Action option and a Feynman quantization of spinor fields in terms of ordinary C numbers,''
Annals Phys. \textbf{11} (1960), 123-168
    
    
\bibitem{Tada:2019rls}
T.~Tada,
``Time development of conformal field theories associated with $L_{1}$ and $L_{-1}$ operators,''
J. Phys. A \textbf{53} (2020) no.25, 255401

 \bibitem{Bueno:2019ajd}
P.~Bueno, J.~M.~Magan and C.~S.~Shahbazi,
``Complexity measures in QFT and constrained geometric actions,''
JHEP \textbf{09} (2021), 200,
[arXiv:1908.03577 [hep-th]].


\bibitem{Erdmenger:2020sup}
J.~Erdmenger, M.~Gerbershagen and A.~L.~Weigel,
``Complexity measures from geometric actions on Virasoro and Kac-Moody orbits,''
JHEP \textbf{11} (2020), 003

 \bibitem{Patramanis}
D.~Patramanis, to appear.


 \bibitem{Agarwal2012QuantumOptics}
Agarwal, Girish S., Quantum Optics, 2012,  Cambridge University Press

\bibitem{DeBoer:2018kvc}
J.~De Boer, J.~J\"arvel\"a and E.~Keski-Vakkuri,
``Aspects of capacity of entanglement,''
Phys. Rev. D \textbf{99} (2019) no.6, 066012

\bibitem{Kawabata:2021hac}
K.~Kawabata, T.~Nishioka, Y.~Okuyama and K.~Watanabe,
``Probing Hawking radiation through capacity of entanglement,''
JHEP \textbf{05} (2021), 062
 
\bibitem{Nandy:2021hmk}
P.~Nandy,
``Capacity of entanglement in local operators,''
JHEP \textbf{07} (2021), 019


  \bibitem{Vidal:2002zz}
G.~Vidal and R.~F.~Werner,
``Computable measure of entanglement,''
Phys. Rev. A \textbf{65} (2002), 032314
[arXiv:quant-ph/0102117 [quant-ph]].

\bibitem{Miyaji:2015woj}
M.~Miyaji, T.~Numasawa, N.~Shiba, T.~Takayanagi and K.~Watanabe,
``Distance between Quantum States and Gauge-Gravity Duality,''
Phys. Rev. Lett. \textbf{115} (2015) no.26, 261602

\bibitem{DiGiulio:2020hlz}
G.~Di Giulio and E.~Tonni,
``Complexity of mixed Gaussian states from Fisher information geometry,''
JHEP \textbf{12} (2020), 101

 \bibitem{Yang:2018nda}
R.~Q.~Yang, Y.~S.~An, C.~Niu, C.~Y.~Zhang and K.~Y.~Kim,
``Principles and symmetries of complexity in quantum field theory,''
Eur. Phys. J. C \textbf{79} (2019) no.2, 109
    
\bibitem{BCzech:2016xec}
   B.~Czech,  L.~Lamprou, S.~McCandlish, B.~Mosk and J.~Sully,
  {\em A Stereoscopic Look into the Bulk},
  JHEP {\bf 07} (2016) 129,
  [arXiv:1604.03110 [hep-th]].
  

 \bibitem{deBoer:2016pqk}
   J.~de Boer, F.~Haehl, M.P.~Heller and R.C.~Myers,
  {\em Entanglement, holography and causal diamonds},
  JHEP {\bf 08} (2016) 162,
  [arXiv:1606.03307 [hep-th]].

\bibitem{Goto:2017olq}
K.~Goto and T.~Takayanagi,
``CFT descriptions of bulk local states in the AdS black holes,''
JHEP \textbf{10} (2017), 153
 
 \bibitem{perm}
J.~M.~Magan,
``Decoherence and microscopic diffusion at the Sachdev-Ye-Kitaev model,''
Phys. Rev. D \textbf{98} (2018) no.2, 026015,
[arXiv:1612.06765 [hep-th]].  

\bibitem{Cardy:2014rqa}
J.~Cardy,
``Thermalization and Revivals after a Quantum Quench in Conformal Field Theory,''
Phys. Rev. Lett. \textbf{112} (2014), 220401

\bibitem{Dyer:2016pou}
E.~Dyer and G.~Gur-Ari,
``2D CFT Partition Functions at Late Times,''
JHEP \textbf{08} (2017), 075


\bibitem{Hu:2020suv}
Q.~Hu, A.~Franco-Rubio and G.~Vidal,
``Emergent universality in critical quantum spin chains: entanglement Virasoro algebra,''
[arXiv:2009.11383 [quant-ph]].

\bibitem{DongshengPC}
P.~Caputa and D.~Ge, to appear. 

  
\end{thebibliography}
\end{document}